\documentclass[prl,twocolumn,twoside,showpacs,floatfix,superscriptaddress]{revtex4-1}

\usepackage{graphicx,color,rotating}
\usepackage{amsmath,amssymb,bm}
\usepackage{dcolumn}

\usepackage{times,txfonts}

\usepackage{multirow}

\usepackage{pifont}
\usepackage{dsfont}

\newcommand{\bra}[1]{\ensuremath{\langle{#1}|\,}}

\newcommand{\ket}[1]{\ensuremath{\,|{#1}\rangle}}

\newcommand{\op}[1]{\ensuremath{#1}}

\newcommand{\SAT}{\ensuremath{\text{N}^2\text{LO}_{\text{SAT}}}}

\newcommand{\elem}[2]{\ensuremath{{}^{#2}\text{#1}}}

\definecolor{FGViolet}{rgb}{0.61,0.32,0.61}
\definecolor{FGDarkBlue}{rgb}{0,0,0.6}
\definecolor{FGBlue}{rgb}{0,0,0.8}
\definecolor{FGLightBlue}{rgb}{0.2, 0.6, 0.8}
\definecolor{FGGreen}{rgb}{0.2,0.7,0.2}
\definecolor{FGLightGreen}{rgb}{0.4,1,0.4}
\definecolor{FGYellow}{rgb}{1,0.95,0}
\definecolor{FGOrange}{rgb}{0.95,0.5,0.1}
\definecolor{FGRed}{rgb}{0.8,0,0}
\definecolor{FGWhite}{rgb}{1,1,1}
\definecolor{FGLightGray}{rgb}{0.8,0.8,0.8}
\definecolor{FGGray}{rgb}{0.5,0.5,0.5}
\definecolor{FGDarkGray}{rgb}{0.3,0.3,0.3}
\definecolor{FGBlack}{rgb}{0,0,0}

\begin{document}

\title{Can Ab Initio Theory Explain the Phenomenon of Parity Inversion in  \elem{Be}{11}?}

\author{Angelo Calci}
\email[]{calci@triumf.ca}
\affiliation{TRIUMF, 4004 Wesbrook Mall, Vancouver, British Columbia, V6T 2A3, Canada}

\author{Petr Navr\'atil}
\email[]{navratil@triumf.ca}
\affiliation{TRIUMF, 4004 Wesbrook Mall, Vancouver, British Columbia, V6T 2A3, Canada}

\author{Robert Roth}
\affiliation{Institut f\"ur Kernphysik, Technische Universit\"at Darmstadt, 64289 Darmstadt, Germany}

\author{J\'er\'emy Dohet-Eraly}
\email[]{Present address: Istituto Nazionale di Fisica Nucleare, Sezione di Pisa, Largo B. Pontecorvo 3, I-56127 Pisa, Italy.}
\affiliation{TRIUMF, 4004 Wesbrook Mall, Vancouver, British Columbia, V6T 2A3, Canada}

\author{Sofia Quaglioni}
\affiliation{Lawrence Livermore National Laboratory, P.O. Box 808, L-414, Livermore, California 94551, USA}

\author{Guillaume Hupin}
\affiliation{Institut de Physique Nucl\'eaire, Universit\'e Paris-Sud, IN2P3/CNRS, F-91406 Orsay Cedex, France}
\affiliation{CEA, DAM, DIF, F-91297 Arpajon, France}

\date{\today}

\begin{abstract}

The weakly bound exotic \elem{Be}{11} nucleus, famous for its ground-state parity inversion and distinct n+\elem{Be}{10} halo structure, is investigated from first principles using chiral two- and three-nucleon forces. An explicit treatment of continuum effects is found to be indispensable. We study the sensitivity of the \elem{Be}{11}  spectrum to the details of the three-nucleon force and demonstrate that only certain chiral interactions are capable of reproducing the parity inversion. With such interactions, the extremely large E1 transition between the bound states is reproduced. We compare our photodisintegration calculations to conflicting experimental data and predict a distinct dip around the $3/2^-_1$ resonance energy. Finally, we predict low-lying $3/2^+$ and $9/2^+$ resonances that are not or not sufficiently measured in experiments.

\end{abstract}

\maketitle


The theoretical understanding of exotic neutron-rich nuclei constitutes a tremendous challenge.
These systems often cannot be explained by mean-field approaches and contradict the regular shell structure. 
The spectrum of \elem{Be}{11} has some very peculiar features. The $1/2^+$ ground state (g.s.) is loosely bound by $502\,\text{keV}$ with respect to the n+\elem{Be}{10} threshold and is separated by only $320\,\text{keV}$ from its parity-inverted $1/2^-$ partner~\cite{TiKe04,KeKw12}, which would be the expected g.s.\ in the standard shell-model picture. 
Such parity inversion, already noticed by Talmi and Unna~\cite{TaUn60} in the early 1960s, is one of the best examples of the disappearance of the $N = 8$ magic number with an increasing neutron to proton ratio. 
The next (n+n+\elem{Be}{9}) breakup threshold appears at $7.31\,\text{MeV}$~\cite{WaAu12}, such that the rich resonance structure at low energies is dominated by the n+\elem{Be}{10} dynamics. 
Peculiar also is the electric-dipole transition strength between the two bound states, which has attracted much attention since its first measurement in 1971~\cite{HaNa71} and was remeasured in 1983~\cite{MiOl83} and 2014~\cite{KwWu14}. It is the strongest known transition between low-lying states, attributed to the halo character of \elem{Be}{11}.

An accurate description of this complex spectrum is anticipated to be sensitive to the details of the nuclear force~\cite{CaRo16}, such that a precise knowledge of the nucleon-nucleon (NN) interaction, desirably obtained from first principles, is crucial. Moreover, the inclusion of three-nucleon (3N) effects has been found to be indispensable for an accurate description of nuclear systems~\cite{PiWi01,NaGu07}. 
The chiral effective field theory constitutes one of the most promising candidates for deriving the nuclear interaction. Formulated by Weinberg~\cite{Weinberg79,Weinberg90,Weinberg91}, it is based on the fundamental symmetries of QCD and uses pions and nucleons as relevant degrees of freedom. Within this theory  NN, 3N and higher many-body interactions arise in a natural hierarchy~\cite{Weinberg79,Weinberg90,Weinberg91,OrRa94,VanKolck94,EpNo02,Epel06}. The details of these interactions depend on the specific choices made during the construction. In particular the way the interactions are constrained to experimental data can have a strong impact~\cite{CaEk16}. 

In this work we tackle the question if \emph{ab initio} calculations can provide an accurate description of the \elem{Be}{11} spectrum and reproduce the experimental ground state.
Pioneering \emph{ab initio} investigations of \elem{Be}{11} did not account for the important effects of 3N forces and were incomplete in the treatment of either long-~\cite{FoNa05} or short-range~\cite{QuNa08,QuNa09} correlations, both of which are crucial to arrive at an accurate description of this system. 

In this Letter, we report the first complete \emph{ab initio} calculations of the \elem{Be}{11} nucleus using the framework of the no-core shell model with continuum (NCSMC)~\cite{BaNa13b,BaNa13c,NaQu16}, which combines the capability to describe the extended n+\elem{Be}{10} configurations of Refs.~\cite{QuNa08,QuNa09} with a  robust treatment of many-body short-range correlations. We adopt a family of chiral interactions in which the NN component is constrained, in a traditional sense, to two-nucleon properties~\cite{EnMa03} and the 3N force is fitted in three- and sometimes four-body systems~\cite{Navr07,GaQu09,RoBi12,RoCa14}. In addition, we also employ a newer chiral interaction, obtained from a simultaneous fit of NN and 3N components to nucleon-nucleon scattering data and selected properties of nuclei as complex as~\elem{O}{25}~\cite{EkJa15,HaEk16,MiBa16}. 

\paragraph{Many-body approach.}

The general idea of the NCSMC is to represent the $A$-nucleon wave function as the generalized cluster expansion
~\cite{BaNa13b,BaNa13c,NaQu16} 
%
\begin{equation}
\ket{\Psi_{A}^{J^{\pi} T}} = \sum_{\lambda} c^{J^{\pi} T}_{\lambda} \ket{A\lambda J^{\pi} T} + \sum_{\nu} \int \mathrm{d}r r^2 \frac{\gamma^{J^{\pi} T}_{\nu}(r)}{r} \op{\mathcal{A}}_{\nu} \ket{\Phi_{\nu r}^{J^{\pi} T}}\,.
\label{eq:NCSMC-basis}
\end{equation}
%
The first term consists of an expansion over the no-core shell model (NCSM) eigenstates of the compound system $\ket{A\lambda J^{\pi} T}$ (here \elem{Be}{11}) indexed by $\lambda$. These states are expanded in a finite harmonic oscillator basis and thus well suited to cover the localized correlations of the $A$-body system, but are inappropriate to describe clustering and scattering properties. The latter properties are addressed by the second term corresponding to an expansion over the antisymmetrized cluster channels $\op{\mathcal{A}}_{\nu} \ket{\Phi_{\nu r}^{J^{\pi} T}}$~\cite{QuNa09}, which describe the two clusters (here n+\elem{Be}{10}) in relative motion. Here $r$ denotes the relative distance of the clusters and $\nu$ is a collective index for the relevant quantum numbers. 
The expansion coefficients  $ c^{J^{\pi} T}_{\lambda}$  and the continuous relative-motion amplitudes $\gamma^{J^{\pi} T}_{\nu}(r)$ are obtained as a solution of the generalized eigenvalue problem derived by representing 
the Schr\"odinger equation in the model space of expansion~\eqref{eq:NCSMC-basis} as detailed in Refs.~\cite{QuNa09,BaNa13c,NaQu16}. The resulting NCSMC equations are solved by the coupled-channel R-matrix method on a Lagrange mesh~\cite{DeBa10,HeRo02,BaGo02}. The resonance energies and widths are deduced from the complex poles of the $S$ matrix, via the $R$-matrix approach extended to complex energies and momenta~\cite{Schn81,DoNa16}.

The inclusion of the 3N force is computationally highly demanding and restricts the current application range of the NCSMC. For nuclei with $A>5$ we rely on an on-the-fly computing of the uncoupled densities discussed in Ref.~\cite{HuLa13}.
The present NCSMC calculations are performed including the first three eigenstates ($0^+,2^+_1,2^+_2$) of the \elem{Be}{10} target, entering the cluster states in~\eqref{eq:NCSMC-basis} and at least the first four negative- and three positive-parity eigenstates of \elem{Be}{11}.
Such eigenstates are obtained within the NCSM, except in the largest model spaces where, to reduce the dimension of the problem, we use the importance-truncated NCSM~\cite{Roth09,RoNa07}. 

\paragraph{Analysis of spectroscopy.}
We start by using an interaction and parameter set established in numerous studies~\cite{HuLa13,LaNa15,RoCa14,BiLa13b,RoLa11} and investigate the convergence with respect to the model-space size $N_{\text{max}}$. We use the traditionally fitted chiral interaction where we choose the cutoff in the 3N regularization to be $\Lambda_{\text{3N}}=400\,\text{MeV}$, indicated by NN+3N(400). 
To accelerate the convergence of the many-body approach the interactions are softened via the similarity renormalization group (SRG)~\cite{Wegn94,BoFu07,SzPe00} as described in Refs.~\cite{RoLa11,RoCa14} (see Supplemental Material  for details~\cite{Supplement}). Note that both the SRG-induced and initial 3N forces are treated explicitly at all steps of the calculations.

%
\begin{figure}[t]
\includegraphics[clip,width=0.9\columnwidth]{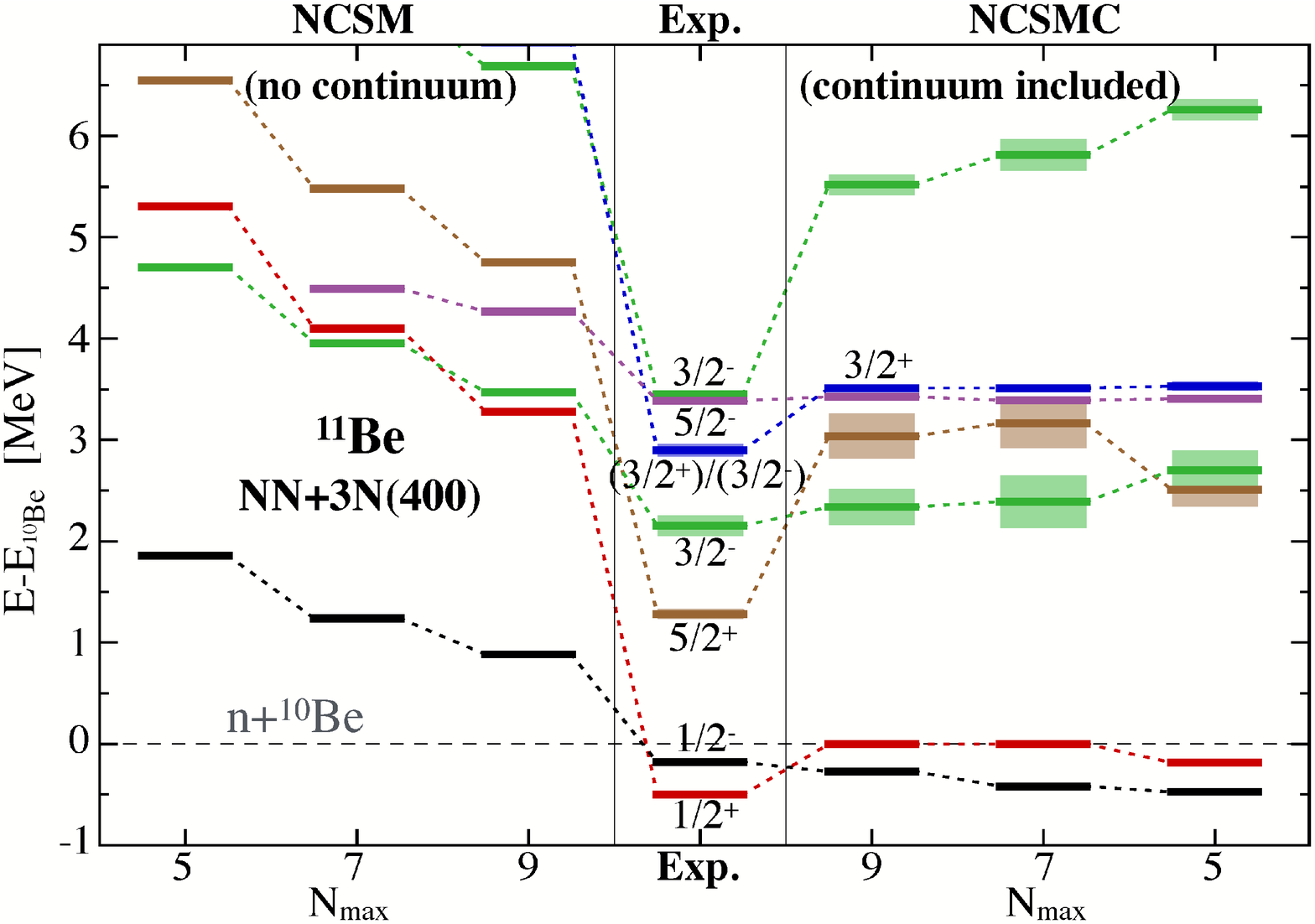}\\[-8pt] 
\caption{(color online) Spectrum of \elem{Be}{11} with respect to the n+\elem{Be}{10} threshold. The NCSM (left) and NCSMC (right) calculations are carried out for different model-space sizes ($N_{\text{max}}=5,7,9$). Light boxes of  experimental and NCSMC spectra indicate resonance widths. Experimental energies taken from~\cite{TiKe04,KeKw12}. See the text and Supplemental Material for details of the calculations~\cite{Supplement}. 
 }
\label{fig:Be11_Spec_Nmax}
\end{figure}
%

Without continuum effects, i.e., using the conventional NCSM a converged \elem{Be}{11} spectrum cannot be obtained within accessible model spaces as demonstrated in Fig.~\ref{fig:Be11_Spec_Nmax}. All states are unbound with respect to the n+\elem{Be}{10} threshold. The positive-parity states converge especially slowly, their excitation energy is too high compared to the experiment. Once continuum effects are taken into account through the inclusion of the n+\elem{Be}{10} cluster states in the model space, the convergence improves drastically, even though the computed threshold energy of n+\elem{Be}{10} is not fully converged, yet. At $N_{\text{max}}=9$, this energy is $-58.4\,\text{MeV}$ and increases by $2.3$ and $6.2\,\text{MeV}$ for the $N_{\text{max}}=7$ and $5$ model spaces, respectively.  The extrapolated value of $-60.9(10)\,\text{MeV}$ is underbound with respect to the experimental energy of $-64.976\,\text{MeV}$~\cite{WaAu12}. 
%
\begin{figure*}[t]

\hspace*{-20pt}\includegraphics[width=0.95\textwidth]{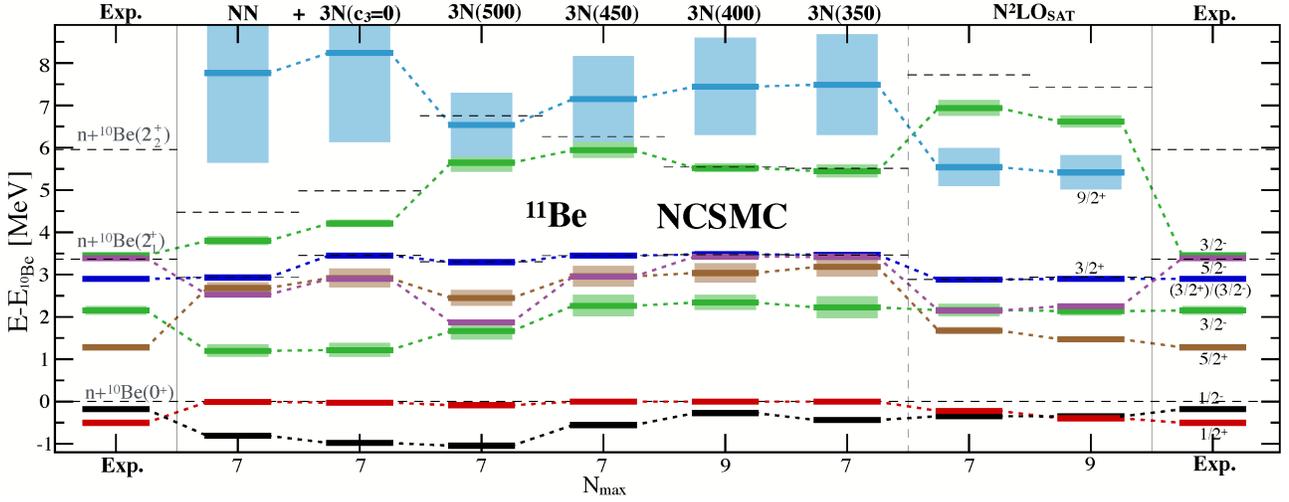}\\[-8pt] 
\caption{(color online) NCSMC spectrum of \elem{Be}{11} with respect to the n+\elem{Be}{10} threshold. 
Dashed black lines indicate the energies of the \elem{Be}{10} states. Light boxes indicate the resonance width. Experimental energies are taken from Refs.~\cite{TiKe04,KeKw12}.
 }
\label{fig:Be11_Spec}
\end{figure*}
%
For the negative parity, the NCSMC achieves an overall quite reasonable description, especially for the three lowest states. On the other hand, the $1/2^+$ state is barely bound and the parity inversion of the bound states is not reproduced. Similarly, the $3/2^-_1$ and $5/2^+$ states are inverted compared to the experiment.   
The $3/2^-_2$ excitation energy is about $2\,\text{MeV}$ larger than the experimental one. 
Other decay channels (and hence cluster states) presently not included may play a role at such high energies.

We first analyze the sensitivity of the spectrum to the 3N interaction in Fig.~\ref{fig:Be11_Spec}.  From left to right we use exclusively the chiral NN interaction (including SRG-induced 3N contributions), the 3N interaction with a $500\,\text{MeV}$ cutoff, where parts of the two-pion exchange contribution are suppressed ($c_{3}=0$), and the full 3N contributions using the cutoffs $\Lambda_{3N}=500, 450, 400,$  and $350\, \text{MeV}$ as introduced in Ref.~\cite{RoCa14}. The illustrated spectra are expected to show a similar convergence pattern as in the case of the NN+3N(400) interaction. The omitted SRG-induced beyond-3N contributions are expected to impact the \elem{Be}{11} spectrum only for the NN+3N(500) interaction, while the remaining spectra are anticipated to be unaffected~\cite{RoLa11,RoBi12,RoCa14}.
We find the two-pion exchange term to cause the dominant 3N effects in the \elem{Be}{11} spectrum. The 3N interactions generally increase the excitation energies of both $3/2^{-}$ resonances, corresponding to the increase in excitation energy of the $2^+$ states in \elem{Be}{10}.
Neither the inversion of the $1/2^+$ and $1/2^-$ states, nor that of the $3/2^{-}_{1}$ and $5/2^+$ states can be explained by the adopted 3N force versions. 
Decreasing the 3N cutoff initially reduces the bound-state splitting, but below $\Lambda_{3N}=400\, \text{MeV}$ the influence of the 3N interaction is too strongly reduced such that the spectra approach the pure NN result. 
On the contrary, the converged spectrum with the simultaneously fitted NN+3N interaction, named  $\SAT$~\cite{EkJa15},  successfully achieves the parity inversions between the $3/2^-_1$ and $5/2^+$ resonances and, albeit marginally, for the bound states. The low-lying spectrum is significantly improved and agrees well with the experiment, presumably due to the more accurate description of long-range properties caused by the fit of the interaction to radii of p-shell nuclei. On the other hand, the strongly overestimated splitting between the $3/2^-_2$ and $5/2^-$ states hints at deficiencies of this interaction, which might originate from a too large splitting of the $p_{1/2}$-$p_{3/2}$ subshells.   

In addition to the resonances observed in the experiment all theoretical spectra predict a low-lying $9/2^+$ resonance suggested in Refs.~\cite{AoYo97,HiSh04}. For the $\SAT$ interaction the resonance energy is close to the one predicted by the Gamow shell model~\cite{FoNa16b}, although our \emph{ab initio} calculations predict a broader width.
Another interesting property is the position of the $3/2^+$ resonance that is strongly influenced by the $2^+_1$ state of \elem{Be}{10}. For all theoretical calculations the energies of these correlated states are almost degenerate, while in the experiment the $2^+_1$ state in \elem{Be}{10} is about $470\,\text{keV}$ above the tentative $3/2^+$ state and coincides with the $3/2^-_2$ and $5/2^-$ resonances. 
  
\paragraph{Nuclear structure and reaction properties.}

\begin{figure}[t]
\includegraphics[clip, width=1.00\columnwidth]{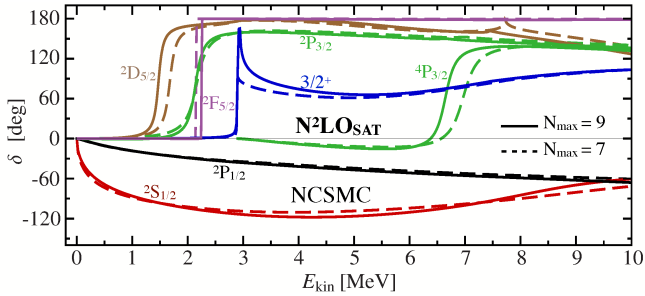}\\[-8pt] 
\caption{(color online) The n+\elem{Be}{10} phase shifts as a function of the kinetic energy in the center-of-mass frame. 
NCSMC phase shifts for the $\SAT$ interaction are compared for two model spaces indicated by $N_{\text{max}}$. 
 }
\label{fig:Be11_PS}
\end{figure}
Except for the two bound states, all the energy levels of Fig.~\ref{fig:Be11_PS} correspond to n+\elem{Be}{10} scattering states. The corresponding phase shifts obtained with the $\SAT$ interaction are presented in Fig.~\ref{fig:Be11_PS} (see Supplemental Material for further details~\cite{Supplement}). The overall proximity of the $N_{\text{max}}=7$ and $9$ results confirms the good convergence with respect to the model space. 
The states observed in \elem{Be}{11} are typically dominated by a single n+\elem{Be}{10} partial wave but the illustrated eigenphase shifts of the $3/2^+$ state consist of a superposition of the $^{4}S_{3/2}$ and $^{2}D_{3/2}$ partial waves. The parity of this resonance is experimentally not uniquely extracted~\cite{TiKe04,KeKw12}, while all \emph{ab initio} calculations concordantly predict it to be positive.
The bound-state energies as well as the resonance energies and widths for different interactions and both many-body approaches are summarized in Tab.~\ref{tab:spectrum}.   
In the case of the NN+3N(400) interaction, however, the fast $3/2^+$ phase shift variation near the n+\elem{Be}{10}$(2^+_1)$ threshold does not correspond to a pole of the scattering matrix, such that this state is not a resonance in the conventional sense and a width could not be extracted reliably.
The theoretical widths tend to overestimate the experimental value, but overall the agreement is reasonable, especially for the $\SAT$ interaction.
Experimentally only an upper bound could be determined for the $5/2^-$ resonance width and the theoretical calculations predict an extremely narrow resonance. 
%
%
\begin{table}[t]
\begin{tabular}{c | cc |cc | cc | cc}
                \hline
                \hline
      \multirow{3}*{$J^{\pi}$}        
     													 & \multicolumn{4}{ c |}{ \footnotesize NCSMC }    															& \multicolumn{2}{ c |}{  \footnotesize NCSMC-pheno} &  \multicolumn{2}{c}{\multirow{2}*{Exp.}}   \\
     													  & \multicolumn{2}{ c |}{  NN+3N(400) }   				& \multicolumn{2}{ c |}{ $\SAT$}	& \multicolumn{2}{ c |}{  $\SAT$ }    \\ 
                                                & $E$       & $\Gamma$    & $E$       & $\Gamma$     &  $E$       & $\Gamma$  & $E$       & $\Gamma$   \\
                \hline
               \hline    
      $1/2^+$  			& -0.001   & -							& -0.40 		&  -   		 	 	& -0.50 & - 					& -0.50 & -\\	
      $1/2^-$ 			&  -0.27    & -								& -0.35 		&  -  	 				& -0.18 & -   				& -0.18 & -\\

      $5/2^+$ 			& 3.03 		& 0.44 					& 1.47 		&   0.12  			& 1.31 	& 0.10  		& 1.28 & 0.1 \\	
      $3/2_{1}^-$ 	& 2.34	 		& 0.35						& 2.14  		&   0.21  			& 2.15 	& 0.19    		& 2.15 & 0.21 \\											                                              
      $3/2^+$ 			& 3.48 		& - 					& 2.90			& 0.014   		& 2.92 	& 0.06   		& 2.898 & 0.122 \\											                                              
      $5/2^-$ 			& 3.43	 	   & 0.001 						& 2.25 	    & 0.0001			& 3.30 	& 0.0002  	& 3.3874 & $<$0.008\\											                                              
      $3/2_{2}^-$ 	& 5.52 		& 0.20						& 6.62 		& 0.29 	     	& 5.72 	& 0.19   		& 3.45 & 0.01 \\	
      $9/2^+$ 			& 7.44 		& 2.30 					& 5.42  		& 0.80 		 	& 5.59 	& 0.62   		& - & - \\	
\hline
\hline
\end{tabular}
\caption{Excitation spectrum of \elem{Be}{11} with respect to the n+\elem{Be}{10} threshold. Energies and widths are in MeV.  The calculations are carried out at $N_{\text{max}}=9$. 
}
\label{tab:spectrum}
\end{table}
%
%

Although the bulk properties of the spectrum are already well described, accurate predictions of observables, such as electric-dipole (E1) transitions, which probe the structure of the nucleus, can be quite sensitive to the energies of the involved states with respect to the threshold. Based on our analysis, the discrepancies between the theoretical and experimental energy spectra can be mostly attributed to deficiencies in the nuclear force. Therefore, it can be beneficial to loosen the first-principles paradigm to remedy the insufficiencies in the nuclear force and provide accurate predictions for complex observables using the structure information of the \emph{ab initio} approach. In the following we use a phenomenology-inspired approach indicated by NCSMC-pheno that has been already applied in Refs.~\cite{RaHu16,DoNa16}. In this approach we adjust the \elem{Be}{10} and \elem{Be}{11} excitation energies of the NCSM eigenstates entering expansion~\eqref{eq:NCSMC-basis} to reproduce the experimental energies of the first low-lying states.
Note that the obtained NCSMC-pheno energies are fitted to the experiment while the theoretical widths, quoted in Tab.~\ref{tab:spectrum}, are predictions. 

\begin{figure}[t]
\includegraphics[width=0.9\columnwidth]{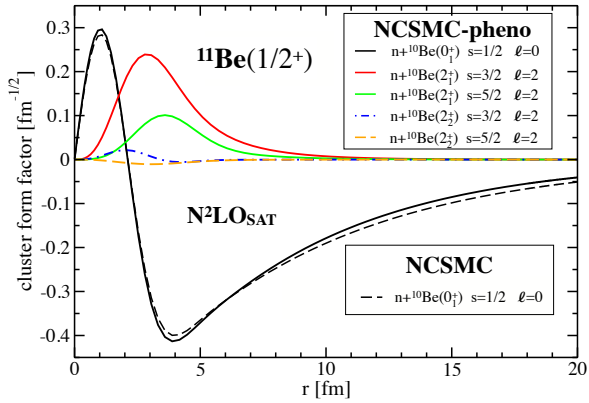}\\[-5pt]
\caption{(color online) Comparison of the cluster form factors with the $\SAT$ interaction at $N_{\text{max}}=9$. Note the coupling between the \elem{Be}{10} target and neutron in the cluster state $\ket{\Phi_{\nu,r}^{J^{\pi} T}}\sim\big[\big(\ket{\elem{Be}{10}: I_{1}^{\pi_{1}}T_1} \,\ket{n:1/2^{+}1/2}\big)^{sT}\, Y_{\ell}(\hat{r})\big]^{J^{\pi} T}$.
}
\label{fig:Be11_formfactor}
\end{figure}
%
An intuitive interpretation of the \elem{Be}{11} g.s.\ wave function, is provided in Fig.~\ref{fig:Be11_formfactor} by the overlap of the full solution for the g.s.\ $\ket{\Psi_{\nu}^{J^{\pi} T}}$ in~\eqref{eq:NCSMC-basis} with the cluster portion $\ket{\Phi_{\nu,r}^{J^{\pi} T}}$ given by $r \cdot \bra{\Phi_{\nu,r}^{J^{\pi} T} } \op{\mathcal{A}}_{\nu} \ket{\Psi_A^{J^{\pi} T}}$. A clearly extended halo structure beyond $20\,\text{fm}$ can be identified for the S wave of the \elem{Be}{10}($0^{+}$)+n relative motion. The phenomenological energy adjustment only slightly influences the asymptotic behavior of the S wave, as seen by comparing the solid and dashed black curves, while other partial waves are even indistinguishable on the plot resolution.
The corresponding spectroscopic factors for the NCSMC-pheno approach, obtained by integrating the squared cluster form factors in Fig.~\ref{fig:Be11_formfactor}, are: $S=0.90$ (S wave) and  S=0.16 (D wave). The S-wave asymptotic normalization coefficient is $0.786\,\text{fm}^{-1/2}$.

%
%
\begin{table}[t]
\begin{tabular}{c   c c c || c }
               \hline
                \hline
                             & NCSM       				& NCSMC       	&  NCSMC-pheno     & Exp.   			\\
                \hline    
     NN+3N(400) 													& 	0.0005						& - 					&     	0.146	 	 		& \multirow{2}*{0.102(2)		~\cite{KwWu14}}   					\\	 

      					$\SAT$							   		   &	0.0005						& 0.127 		&	0.117		          \\  
     \hline
      \hline	
\end{tabular}
\caption{Reduced transition probability $B$(E1:\,$1/2^- \rightarrow 1/2^+$) between \elem{Be}{11} bound states in $\text{e}^2\text{fm}^2$.   
}
\label{tab:E1}
\end{table}
%
%

The B(E1) transitions are summarized in Tab.~\ref{tab:E1}.
Calculations without continuum effects predict the wrong g.s.\ and underestimate the E1 strength by several orders of magnitude. For the NCSMC calculations with the NN+3N(400) interaction the $1/2^+$ state is very weakly bound leading to an unrealistic E1 transition. The $\SAT$ interaction successfully reproduces the strong E1 transition, albeit the latest measurement~\cite{KwWu14} is slightly overestimated, even after the phenomenological energy adjustment. 
There might be small effects arising from a formally necessary SRG evolution of the transition operator.
Works along these lines for \elem{He}{4} suggest a slight reduction of the dipole strength~\cite{ScQu14,ScQu15}. A similar effect would bring the calculated E1 transition in better agreement with the experiment~\cite{KwWu14}.  

%
\begin{figure}[t]
\vspace{-4pt}
\hspace{-10pt}
\includegraphics[width=0.9\columnwidth]{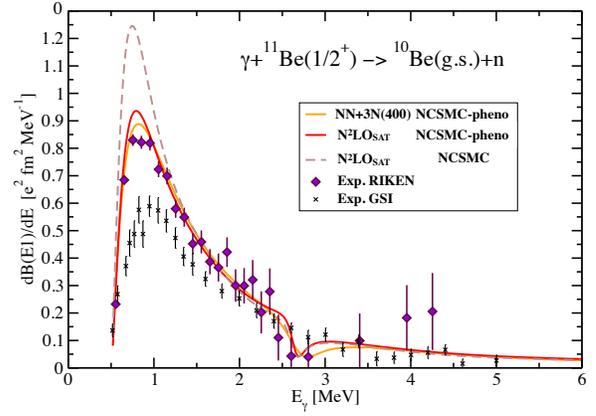}\\[-10pt]
\caption{(color online)
Dipole strength distribution $dB(E1)/dE$ of the photodisintegration process as a
function of the photon energy. Theoretical dipole strength distributions for two chiral interactions with (solid line) and without (dashed line) the phenomenological energy adjustment are compared to the experimental measurements at GSI~\cite{PaAd03,AuNa13} (black dots) and RIKEN~\cite{NaSh94,FuNa04,AuNa13} (violet dots).
}
\label{fig:Be11_photodis}
\end{figure}

Finally we study the photodisintegration of the \elem{Be}{11} g.s.\ into n+\elem{Be}{10} in Fig.~\ref{fig:Be11_photodis}.
This is proportional to dipole strength distribution $dB(E1)/dE$.     
In all approaches, a peak of nonresonant nature (see Fig.~\ref{fig:Be11_PS}) is present at about $800\,\text{keV}$ above the n+\elem{Be}{10} threshold, particularly pronounced in the $3/2^-$ partial wave.
The strong peak for the NCSMC with the $\SAT$ interaction is caused by the slightly extended S wave tail in Fig.~\ref{fig:Be11_formfactor} and hence the underestimated binding energy of the $1/2^+$ state. 
The theoretical predictions are compared to indirect measurements of the photodissociation process extracted from the scattering experiments of \elem{Be}{11} on lead~\cite{NaSh94,FuNa04,AuNa13} and carbon~\cite{PaAd03,AuNa13} targets. Our phenomenological adjusted calculations show good agreement with the RIKEN data~\cite{NaSh94,FuNa04,AuNa13}. Based on the analysis in Refs.~\cite{ScQu14,ScQu15}, it is doubtful that the missing SRG evolution of the E1-transition operator could explain the $\sim$30\% discrepancy with the GSI data~\cite{PaAd03,AuNa13}.
A dip in the dipole strength distribution is present at about $2.7\,\text{MeV}$, due to the $3/2^-_1$ resonance.
At this energy, the E1 matrix element between the \elem{Be}{11} g.s.\ wave function and the $3/2^-$ partial wave of the n+\elem{Be}{10} scattering wave function changes its sign. Because of large uncertainties, the experimental data neither confirm nor exclude such a dip. A similar feature, but much less pronounced, can be noticed in microscopic cluster calculations~\cite{Desc97} (see Supplemental Material for details~\cite{Supplement}).

\paragraph{Conclusions.}
We have demonstrated that the inclusion of continuum effects is crucial for a description of the \elem{Be}{11} system and, further that the spectrum is extremely sensitive to the details of the nuclear NN+3N interactions and constitutes an important benchmark for future forces. In particular, the parity inversion of the bound states could be achieved only by the $\SAT$ interaction that provides accurate predictions of nuclear radii and matter saturation properties~\cite{EkJa15,HaEk16}.  
An interesting related endeavor is the investigation of the mirror system p+\elem{C}{10}. New experiments have been performed for the elastic scattering process~\cite{IRIS} that will be analyzed with the NCSMC.   
        
\begin{acknowledgments}
This work was prepared in part by LLNL under Contract No. DE-AC52-07NA27344 and supported by the U.S. Department of Energy, Office of Science, Office of Nuclear Physics, under Work Proposal No. SCW1158, by the NSERC Grant No. SAPIN-2016-00033, and by the Deutsche Forschungsgemeinschaft through SFB 1245. TRIUMF receives federal funding via a contribution agreement with the National Research Council of Canada. Computing support came from the LLNL institutional Computing Grand Challenge Program, from an INCITE Award on the Titan supercomputer of the Oak Ridge Leadership Computing Facility (OLCF) at ORNL, the LOEWE-CSC Frankfurt, the computing center of the TU Darmstadt (LICHTENBERG), and from Calcul Quebec and Compute Canada. 
\end{acknowledgments}


\begin{thebibliography}{63}%
\makeatletter
\providecommand \@ifxundefined [1]{%
 \@ifx{#1\undefined}
}%
\providecommand \@ifnum [1]{%
 \ifnum #1\expandafter \@firstoftwo
 \else \expandafter \@secondoftwo
 \fi
}%
\providecommand \@ifx [1]{%
 \ifx #1\expandafter \@firstoftwo
 \else \expandafter \@secondoftwo
 \fi
}%
\providecommand \natexlab [1]{#1}%
\providecommand \enquote  [1]{``#1''}%
\providecommand \bibnamefont  [1]{#1}%
\providecommand \bibfnamefont [1]{#1}%
\providecommand \citenamefont [1]{#1}%
\providecommand \href@noop [0]{\@secondoftwo}%
\providecommand \href [0]{\begingroup \@sanitize@url \@href}%
\providecommand \@href[1]{\@@startlink{#1}\@@href}%
\providecommand \@@href[1]{\endgroup#1\@@endlink}%
\providecommand \@sanitize@url [0]{\catcode `\\12\catcode `\$12\catcode
  `\&12\catcode `\#12\catcode `\^12\catcode `\_12\catcode `\%12\relax}%
\providecommand \@@startlink[1]{}%
\providecommand \@@endlink[0]{}%
\providecommand \url  [0]{\begingroup\@sanitize@url \@url }%
\providecommand \@url [1]{\endgroup\@href {#1}{\urlprefix }}%
\providecommand \urlprefix  [0]{URL }%
\providecommand \Eprint [0]{\href }%
\providecommand \doibase [0]{http://dx.doi.org/}%
\providecommand \selectlanguage [0]{\@gobble}%
\providecommand \bibinfo  [0]{\@secondoftwo}%
\providecommand \bibfield  [0]{\@secondoftwo}%
\providecommand \translation [1]{[#1]}%
\providecommand \BibitemOpen [0]{}%
\providecommand \bibitemStop [0]{}%
\providecommand \bibitemNoStop [0]{.\EOS\space}%
\providecommand \EOS [0]{\spacefactor3000\relax}%
\providecommand \BibitemShut  [1]{\csname bibitem#1\endcsname}%
\let\auto@bib@innerbib\@empty
\bibitem [{\citenamefont {Tilley}\ \emph {et~al.}(2004)\citenamefont {Tilley},
  \citenamefont {Kelley}, \citenamefont {Godwin}, \citenamefont {Millener},
  \citenamefont {Purcell}, \citenamefont {Sheu},\ and\ \citenamefont
  {Weller}}]{TiKe04}%
  \BibitemOpen
  \bibfield  {author} {\bibinfo {author} {\bibfnamefont {D.}~\bibnamefont
  {Tilley}}, \bibinfo {author} {\bibfnamefont {J.}~\bibnamefont {Kelley}},
  \bibinfo {author} {\bibfnamefont {J.}~\bibnamefont {Godwin}}, \bibinfo
  {author} {\bibfnamefont {D.}~\bibnamefont {Millener}}, \bibinfo {author}
  {\bibfnamefont {J.}~\bibnamefont {Purcell}}, \bibinfo {author} {\bibfnamefont
  {C.}~\bibnamefont {Sheu}}, \ and\ \bibinfo {author} {\bibfnamefont
  {H.}~\bibnamefont {Weller}},\ }\href {\doibase
  http://dx.doi.org/10.1016/j.nuclphysa.2004.09.059} {\bibfield  {journal}
  {\bibinfo  {journal} {Nucl. Phys. A}\ }\textbf {\bibinfo {volume} {745}},\
  \bibinfo {pages} {155 } (\bibinfo {year} {2004})}\BibitemShut {NoStop}%
\bibitem [{\citenamefont {Kelley}\ \emph {et~al.}(2012)\citenamefont {Kelley},
  \citenamefont {Kwan}, \citenamefont {Purcell}, \citenamefont {Sheu},\ and\
  \citenamefont {Weller}}]{KeKw12}%
  \BibitemOpen
  \bibfield  {author} {\bibinfo {author} {\bibfnamefont {J.}~\bibnamefont
  {Kelley}}, \bibinfo {author} {\bibfnamefont {E.}~\bibnamefont {Kwan}},
  \bibinfo {author} {\bibfnamefont {J.}~\bibnamefont {Purcell}}, \bibinfo
  {author} {\bibfnamefont {C.}~\bibnamefont {Sheu}}, \ and\ \bibinfo {author}
  {\bibfnamefont {H.}~\bibnamefont {Weller}},\ }\href {\doibase
  http://dx.doi.org/10.1016/j.nuclphysa.2012.01.010} {\bibfield  {journal}
  {\bibinfo  {journal} {Nucl. Phys. A}\ }\textbf {\bibinfo {volume} {880}},\
  \bibinfo {pages} {88 } (\bibinfo {year} {2012})}\BibitemShut {NoStop}%
\bibitem [{\citenamefont {Talmi}\ and\ \citenamefont {Unna}(1960)}]{TaUn60}%
  \BibitemOpen
  \bibfield  {author} {\bibinfo {author} {\bibfnamefont {I.}~\bibnamefont
  {Talmi}}\ and\ \bibinfo {author} {\bibfnamefont {I.}~\bibnamefont {Unna}},\
  }\href {\doibase 10.1103/PhysRevLett.4.469} {\bibfield  {journal} {\bibinfo
  {journal} {Phys. Rev. Lett.}\ }\textbf {\bibinfo {volume} {4}},\ \bibinfo
  {pages} {469} (\bibinfo {year} {1960})}\BibitemShut {NoStop}%
\bibitem [{\citenamefont {Wang}\ \emph {et~al.}(2012)\citenamefont {Wang},
  \citenamefont {Audi}, \citenamefont {Wapstra}, \citenamefont {Kondev},
  \citenamefont {MacCormick}, \citenamefont {Xu},\ and\ \citenamefont
  {Pfeiffer}}]{WaAu12}%
  \BibitemOpen
  \bibfield  {author} {\bibinfo {author} {\bibfnamefont {M.}~\bibnamefont
  {Wang}}, \bibinfo {author} {\bibfnamefont {G.}~\bibnamefont {Audi}}, \bibinfo
  {author} {\bibfnamefont {A.~H.}\ \bibnamefont {Wapstra}}, \bibinfo {author}
  {\bibfnamefont {F.~G.}\ \bibnamefont {Kondev}}, \bibinfo {author}
  {\bibfnamefont {M.}~\bibnamefont {MacCormick}}, \bibinfo {author}
  {\bibfnamefont {X.}~\bibnamefont {Xu}}, \ and\ \bibinfo {author}
  {\bibfnamefont {B.}~\bibnamefont {Pfeiffer}},\ }\href
  {http://stacks.iop.org/1674-1137/36/i=12/a=003} {\bibfield  {journal}
  {\bibinfo  {journal} {Chin. Phys. C}\ }\textbf {\bibinfo {volume} {36}},\
  \bibinfo {pages} {1603} (\bibinfo {year} {2012})}\BibitemShut {NoStop}%
\bibitem [{\citenamefont {Hanna}\ \emph {et~al.}(1971)\citenamefont {Hanna},
  \citenamefont {Nagatani}, \citenamefont {Harris},\ and\ \citenamefont
  {Olness}}]{HaNa71}%
  \BibitemOpen
  \bibfield  {author} {\bibinfo {author} {\bibfnamefont {S.~S.}\ \bibnamefont
  {Hanna}}, \bibinfo {author} {\bibfnamefont {K.}~\bibnamefont {Nagatani}},
  \bibinfo {author} {\bibfnamefont {W.~R.}\ \bibnamefont {Harris}}, \ and\
  \bibinfo {author} {\bibfnamefont {J.~W.}\ \bibnamefont {Olness}},\ }\href
  {\doibase 10.1103/PhysRevC.3.2198} {\bibfield  {journal} {\bibinfo  {journal}
  {Phys. Rev. C}\ }\textbf {\bibinfo {volume} {3}},\ \bibinfo {pages} {2198}
  (\bibinfo {year} {1971})}\BibitemShut {NoStop}%
\bibitem [{\citenamefont {Millener}\ \emph {et~al.}(1983)\citenamefont
  {Millener}, \citenamefont {Olness}, \citenamefont {Warburton},\ and\
  \citenamefont {Hanna}}]{MiOl83}%
  \BibitemOpen
  \bibfield  {author} {\bibinfo {author} {\bibfnamefont {D.~J.}\ \bibnamefont
  {Millener}}, \bibinfo {author} {\bibfnamefont {J.~W.}\ \bibnamefont
  {Olness}}, \bibinfo {author} {\bibfnamefont {E.~K.}\ \bibnamefont
  {Warburton}}, \ and\ \bibinfo {author} {\bibfnamefont {S.~S.}\ \bibnamefont
  {Hanna}},\ }\href {\doibase 10.1103/PhysRevC.28.497} {\bibfield  {journal}
  {\bibinfo  {journal} {Phys. Rev. C}\ }\textbf {\bibinfo {volume} {28}},\
  \bibinfo {pages} {497} (\bibinfo {year} {1983})}\BibitemShut {NoStop}%
\bibitem [{\citenamefont {Kwan}\ \emph {et~al.}(2014)\citenamefont {Kwan},
  \citenamefont {Wu}, \citenamefont {Summers}, \citenamefont {Hackman},
  \citenamefont {Drake}, \citenamefont {Andreoiu}, \citenamefont {Ashley},
  \citenamefont {Ball}, \citenamefont {Bender}, \citenamefont {Boston},
  \citenamefont {Boston}, \citenamefont {Chester}, \citenamefont {Close},
  \citenamefont {Cline}, \citenamefont {Cross}, \citenamefont {Dunlop},
  \citenamefont {Finlay}, \citenamefont {Garnsworthy}, \citenamefont {Hayes},
  \citenamefont {Laffoley}, \citenamefont {Nano}, \citenamefont {Navrátil},
  \citenamefont {Pearson}, \citenamefont {Pore}, \citenamefont {Quaglioni},
  \citenamefont {Svensson}, \citenamefont {Starosta}, \citenamefont {Thompson},
  \citenamefont {Voss}, \citenamefont {Williams},\ and\ \citenamefont
  {Wang}}]{KwWu14}%
  \BibitemOpen
  \bibfield  {author} {\bibinfo {author} {\bibfnamefont {E.}~\bibnamefont
  {Kwan}}, \bibinfo {author} {\bibfnamefont {C.}~\bibnamefont {Wu}}, \bibinfo
  {author} {\bibfnamefont {N.}~\bibnamefont {Summers}}, \bibinfo {author}
  {\bibfnamefont {G.}~\bibnamefont {Hackman}}, \bibinfo {author} {\bibfnamefont
  {T.}~\bibnamefont {Drake}}, \bibinfo {author} {\bibfnamefont
  {C.}~\bibnamefont {Andreoiu}}, \bibinfo {author} {\bibfnamefont
  {R.}~\bibnamefont {Ashley}}, \bibinfo {author} {\bibfnamefont
  {G.}~\bibnamefont {Ball}}, \bibinfo {author} {\bibfnamefont {P.}~\bibnamefont
  {Bender}}, \bibinfo {author} {\bibfnamefont {A.}~\bibnamefont {Boston}},
  \bibinfo {author} {\bibfnamefont {H.}~\bibnamefont {Boston}}, \bibinfo
  {author} {\bibfnamefont {A.}~\bibnamefont {Chester}}, \bibinfo {author}
  {\bibfnamefont {A.}~\bibnamefont {Close}}, \bibinfo {author} {\bibfnamefont
  {D.}~\bibnamefont {Cline}}, \bibinfo {author} {\bibfnamefont
  {D.}~\bibnamefont {Cross}}, \bibinfo {author} {\bibfnamefont
  {R.}~\bibnamefont {Dunlop}}, \bibinfo {author} {\bibfnamefont
  {A.}~\bibnamefont {Finlay}}, \bibinfo {author} {\bibfnamefont
  {A.}~\bibnamefont {Garnsworthy}}, \bibinfo {author} {\bibfnamefont
  {A.}~\bibnamefont {Hayes}}, \bibinfo {author} {\bibfnamefont
  {A.}~\bibnamefont {Laffoley}}, \bibinfo {author} {\bibfnamefont
  {T.}~\bibnamefont {Nano}}, \bibinfo {author} {\bibfnamefont {P.}~\bibnamefont
  {Navrátil}}, \bibinfo {author} {\bibfnamefont {C.}~\bibnamefont {Pearson}},
  \bibinfo {author} {\bibfnamefont {J.}~\bibnamefont {Pore}}, \bibinfo {author}
  {\bibfnamefont {S.}~\bibnamefont {Quaglioni}}, \bibinfo {author}
  {\bibfnamefont {C.}~\bibnamefont {Svensson}}, \bibinfo {author}
  {\bibfnamefont {K.}~\bibnamefont {Starosta}}, \bibinfo {author}
  {\bibfnamefont {I.}~\bibnamefont {Thompson}}, \bibinfo {author}
  {\bibfnamefont {P.}~\bibnamefont {Voss}}, \bibinfo {author} {\bibfnamefont
  {S.}~\bibnamefont {Williams}}, \ and\ \bibinfo {author} {\bibfnamefont
  {Z.}~\bibnamefont {Wang}},\ }\href {\doibase
  http://dx.doi.org/10.1016/j.physletb.2014.03.049} {\bibfield  {journal}
  {\bibinfo  {journal} {Phys. Lett. B}\ }\textbf {\bibinfo {volume} {732}},\
  \bibinfo {pages} {210 } (\bibinfo {year} {2014})}\BibitemShut {NoStop}%
\bibitem [{\citenamefont {Calci}\ and\ \citenamefont {Roth}(2016)}]{CaRo16}%
  \BibitemOpen
  \bibfield  {author} {\bibinfo {author} {\bibfnamefont {A.}~\bibnamefont
  {Calci}}\ and\ \bibinfo {author} {\bibfnamefont {R.}~\bibnamefont {Roth}},\
  }\href {\doibase 10.1103/PhysRevC.94.014322} {\bibfield  {journal} {\bibinfo
  {journal} {Phys. Rev. C}\ }\textbf {\bibinfo {volume} {94}},\ \bibinfo
  {pages} {014322} (\bibinfo {year} {2016})}\BibitemShut {NoStop}%
\bibitem [{\citenamefont {Pieper}\ and\ \citenamefont
  {Wiringa}(2001)}]{PiWi01}%
  \BibitemOpen
  \bibfield  {author} {\bibinfo {author} {\bibfnamefont {S.~C.}\ \bibnamefont
  {Pieper}}\ and\ \bibinfo {author} {\bibfnamefont {R.~B.}\ \bibnamefont
  {Wiringa}},\ }\href@noop {} {\bibfield  {journal} {\bibinfo  {journal} {Annu.
  Rev. Nucl. Part. Sci.}\ }\textbf {\bibinfo {volume} {51}},\ \bibinfo {pages}
  {53} (\bibinfo {year} {2001})}\BibitemShut {NoStop}%
\bibitem [{\citenamefont {Navr\'atil}\ \emph {et~al.}(2007)\citenamefont
  {Navr\'atil}, \citenamefont {Gueorguiev}, \citenamefont {Vary}, \citenamefont
  {Ormand},\ and\ \citenamefont {Nogga}}]{NaGu07}%
  \BibitemOpen
  \bibfield  {author} {\bibinfo {author} {\bibfnamefont {P.}~\bibnamefont
  {Navr\'atil}}, \bibinfo {author} {\bibfnamefont {V.~G.}\ \bibnamefont
  {Gueorguiev}}, \bibinfo {author} {\bibfnamefont {J.~P.}\ \bibnamefont
  {Vary}}, \bibinfo {author} {\bibfnamefont {W.~E.}\ \bibnamefont {Ormand}}, \
  and\ \bibinfo {author} {\bibfnamefont {A.}~\bibnamefont {Nogga}},\
  }\href@noop {} {\bibfield  {journal} {\bibinfo  {journal} {Phys. Rev. Lett.}\
  }\textbf {\bibinfo {volume} {99}},\ \bibinfo {pages} {042501} (\bibinfo
  {year} {2007})}\BibitemShut {NoStop}%
\bibitem [{\citenamefont {Weinberg}(1979)}]{Weinberg79}%
  \BibitemOpen
  \bibfield  {author} {\bibinfo {author} {\bibfnamefont {S.}~\bibnamefont
  {Weinberg}},\ }\href {\doibase
  http://dx.doi.org/10.1016/0378-4371(79)90223-1} {\bibfield  {journal}
  {\bibinfo  {journal} {Physica (Amsterdam) A}\ }\textbf {\bibinfo {volume}
  {96}},\ \bibinfo {pages} {327} (\bibinfo {year} {1979})}\BibitemShut
  {NoStop}%
\bibitem [{\citenamefont {Weinberg}(1990)}]{Weinberg90}%
  \BibitemOpen
  \bibfield  {author} {\bibinfo {author} {\bibfnamefont {S.}~\bibnamefont
  {Weinberg}},\ }\href {\doibase
  http://dx.doi.org/10.1016/0370-2693(90)90938-3} {\bibfield  {journal}
  {\bibinfo  {journal} {Phys. Lett. B}\ }\textbf {\bibinfo {volume} {251}},\
  \bibinfo {pages} {288} (\bibinfo {year} {1990})}\BibitemShut {NoStop}%
\bibitem [{\citenamefont {Weinberg}(1991)}]{Weinberg91}%
  \BibitemOpen
  \bibfield  {author} {\bibinfo {author} {\bibfnamefont {S.}~\bibnamefont
  {Weinberg}},\ }\href {\doibase
  http://dx.doi.org/10.1016/0550-3213(91)90231-L} {\bibfield  {journal}
  {\bibinfo  {journal} {Nucl. Phys. B}\ }\textbf {\bibinfo {volume} {363}},\
  \bibinfo {pages} {3} (\bibinfo {year} {1991})}\BibitemShut {NoStop}%
\bibitem [{\citenamefont {Ord\'o\~nez}\ \emph {et~al.}(1994)\citenamefont
  {Ord\'o\~nez}, \citenamefont {Ray},\ and\ \citenamefont {van
  Kolck}}]{OrRa94}%
  \BibitemOpen
  \bibfield  {author} {\bibinfo {author} {\bibfnamefont {C.}~\bibnamefont
  {Ord\'o\~nez}}, \bibinfo {author} {\bibfnamefont {L.}~\bibnamefont {Ray}}, \
  and\ \bibinfo {author} {\bibfnamefont {U.}~\bibnamefont {van Kolck}},\ }\href
  {\doibase 10.1103/PhysRevLett.72.1982} {\bibfield  {journal} {\bibinfo
  {journal} {Phys. Rev. Lett.}\ }\textbf {\bibinfo {volume} {72}},\ \bibinfo
  {pages} {1982} (\bibinfo {year} {1994})}\BibitemShut {NoStop}%
\bibitem [{\citenamefont {van Kolck}(1994)}]{VanKolck94}%
  \BibitemOpen
  \bibfield  {author} {\bibinfo {author} {\bibfnamefont {U.}~\bibnamefont {van
  Kolck}},\ }\href {\doibase 10.1103/PhysRevC.49.2932} {\bibfield  {journal}
  {\bibinfo  {journal} {Phys. Rev. C}\ }\textbf {\bibinfo {volume} {49}},\
  \bibinfo {pages} {2932} (\bibinfo {year} {1994})}\BibitemShut {NoStop}%
\bibitem [{\citenamefont {Epelbaum}\ \emph {et~al.}(2002)\citenamefont
  {Epelbaum}, \citenamefont {Nogga}, \citenamefont {Gl\"ockle}, \citenamefont
  {Kamada}, \citenamefont {Mei\ss{}ner},\ and\ \citenamefont
  {Wita\l{}a}}]{EpNo02}%
  \BibitemOpen
  \bibfield  {author} {\bibinfo {author} {\bibfnamefont {E.}~\bibnamefont
  {Epelbaum}}, \bibinfo {author} {\bibfnamefont {A.}~\bibnamefont {Nogga}},
  \bibinfo {author} {\bibfnamefont {W.}~\bibnamefont {Gl\"ockle}}, \bibinfo
  {author} {\bibfnamefont {H.}~\bibnamefont {Kamada}}, \bibinfo {author}
  {\bibfnamefont {U.-G.}\ \bibnamefont {Mei\ss{}ner}}, \ and\ \bibinfo {author}
  {\bibfnamefont {H.}~\bibnamefont {Wita\l{}a}},\ }\href@noop {} {\bibfield
  {journal} {\bibinfo  {journal} {Phys. Rev. C}\ }\textbf {\bibinfo {volume}
  {66}},\ \bibinfo {pages} {064001} (\bibinfo {year} {2002})}\BibitemShut
  {NoStop}%
\bibitem [{\citenamefont {Epelbaum}(2006)}]{Epel06}%
  \BibitemOpen
  \bibfield  {author} {\bibinfo {author} {\bibfnamefont {E.}~\bibnamefont
  {Epelbaum}},\ }\href {\doibase http://dx.doi.org/10.1016/j.ppnp.2005.09.002}
  {\bibfield  {journal} {\bibinfo  {journal} {Prog. Part. Nucl.}\ }\textbf
  {\bibinfo {volume} {57}},\ \bibinfo {pages} {654} (\bibinfo {year}
  {2006})}\BibitemShut {NoStop}%
\bibitem [{\citenamefont {Carlsson}\ \emph {et~al.}(2016)\citenamefont
  {Carlsson}, \citenamefont {Ekstr\"om}, \citenamefont {Forss\'en},
  \citenamefont {Str\"omberg}, \citenamefont {Jansen}, \citenamefont {Lilja},
  \citenamefont {Lindby}, \citenamefont {Mattsson},\ and\ \citenamefont
  {Wendt}}]{CaEk16}%
  \BibitemOpen
  \bibfield  {author} {\bibinfo {author} {\bibfnamefont {B.~D.}\ \bibnamefont
  {Carlsson}}, \bibinfo {author} {\bibfnamefont {A.}~\bibnamefont {Ekstr\"om}},
  \bibinfo {author} {\bibfnamefont {C.}~\bibnamefont {Forss\'en}}, \bibinfo
  {author} {\bibfnamefont {D.~F.}\ \bibnamefont {Str\"omberg}}, \bibinfo
  {author} {\bibfnamefont {G.~R.}\ \bibnamefont {Jansen}}, \bibinfo {author}
  {\bibfnamefont {O.}~\bibnamefont {Lilja}}, \bibinfo {author} {\bibfnamefont
  {M.}~\bibnamefont {Lindby}}, \bibinfo {author} {\bibfnamefont {B.~A.}\
  \bibnamefont {Mattsson}}, \ and\ \bibinfo {author} {\bibfnamefont {K.~A.}\
  \bibnamefont {Wendt}},\ }\href {\doibase 10.1103/PhysRevX.6.011019}
  {\bibfield  {journal} {\bibinfo  {journal} {Phys. Rev. X}\ }\textbf {\bibinfo
  {volume} {6}},\ \bibinfo {pages} {011019} (\bibinfo {year}
  {2016})}\BibitemShut {NoStop}%
\bibitem [{\citenamefont {Forss\'en}\ \emph {et~al.}(2005)\citenamefont
  {Forss\'en}, \citenamefont {Navr\'atil}, \citenamefont {Ormand},\ and\
  \citenamefont {Caurier}}]{FoNa05}%
  \BibitemOpen
  \bibfield  {author} {\bibinfo {author} {\bibfnamefont {C.}~\bibnamefont
  {Forss\'en}}, \bibinfo {author} {\bibfnamefont {P.}~\bibnamefont
  {Navr\'atil}}, \bibinfo {author} {\bibfnamefont {W.~E.}\ \bibnamefont
  {Ormand}}, \ and\ \bibinfo {author} {\bibfnamefont {E.}~\bibnamefont
  {Caurier}},\ }\href {\doibase 10.1103/PhysRevC.71.044312} {\bibfield
  {journal} {\bibinfo  {journal} {Phys. Rev. C}\ }\textbf {\bibinfo {volume}
  {71}},\ \bibinfo {pages} {044312} (\bibinfo {year} {2005})}\BibitemShut
  {NoStop}%
\bibitem [{\citenamefont {Quaglioni}\ and\ \citenamefont
  {Navr\'atil}(2008)}]{QuNa08}%
  \BibitemOpen
  \bibfield  {author} {\bibinfo {author} {\bibfnamefont {S.}~\bibnamefont
  {Quaglioni}}\ and\ \bibinfo {author} {\bibfnamefont {P.}~\bibnamefont
  {Navr\'atil}},\ }\href {http://link.aps.org/abstract/PRL/v101/e092501}
  {\bibfield  {journal} {\bibinfo  {journal} {Phys. Rev. Lett.}\ }\textbf
  {\bibinfo {volume} {101}},\ \bibinfo {pages} {092501} (\bibinfo {year}
  {2008})}\BibitemShut {NoStop}%
\bibitem [{\citenamefont {Quaglioni}\ and\ \citenamefont
  {Navr\'atil}(2009)}]{QuNa09}%
  \BibitemOpen
  \bibfield  {author} {\bibinfo {author} {\bibfnamefont {S.}~\bibnamefont
  {Quaglioni}}\ and\ \bibinfo {author} {\bibfnamefont {P.}~\bibnamefont
  {Navr\'atil}},\ }\href {\doibase 10.1103/PhysRevC.79.044606} {\bibfield
  {journal} {\bibinfo  {journal} {Phys. Rev. C}\ }\textbf {\bibinfo {volume}
  {79}},\ \bibinfo {pages} {044606} (\bibinfo {year} {2009})}\BibitemShut
  {NoStop}%
\bibitem [{\citenamefont {Baroni}\ \emph
  {et~al.}(2013{\natexlab{a}})\citenamefont {Baroni}, \citenamefont
  {Navr\'atil},\ and\ \citenamefont {Quaglioni}}]{BaNa13b}%
  \BibitemOpen
  \bibfield  {author} {\bibinfo {author} {\bibfnamefont {S.}~\bibnamefont
  {Baroni}}, \bibinfo {author} {\bibfnamefont {P.}~\bibnamefont {Navr\'atil}},
  \ and\ \bibinfo {author} {\bibfnamefont {S.}~\bibnamefont {Quaglioni}},\
  }\href {\doibase 10.1103/PhysRevLett.110.022505} {\bibfield  {journal}
  {\bibinfo  {journal} {Phys. Rev. Lett.}\ }\textbf {\bibinfo {volume} {110}},\
  \bibinfo {pages} {022505} (\bibinfo {year} {2013}{\natexlab{a}})}\BibitemShut
  {NoStop}%
\bibitem [{\citenamefont {Baroni}\ \emph
  {et~al.}(2013{\natexlab{b}})\citenamefont {Baroni}, \citenamefont
  {Navr\'atil},\ and\ \citenamefont {Quaglioni}}]{BaNa13c}%
  \BibitemOpen
  \bibfield  {author} {\bibinfo {author} {\bibfnamefont {S.}~\bibnamefont
  {Baroni}}, \bibinfo {author} {\bibfnamefont {P.}~\bibnamefont {Navr\'atil}},
  \ and\ \bibinfo {author} {\bibfnamefont {S.}~\bibnamefont {Quaglioni}},\
  }\href {\doibase 10.1103/PhysRevC.87.034326} {\bibfield  {journal} {\bibinfo
  {journal} {Phys. Rev. C}\ }\textbf {\bibinfo {volume} {87}},\ \bibinfo
  {pages} {034326} (\bibinfo {year} {2013}{\natexlab{b}})}\BibitemShut
  {NoStop}%
\bibitem [{\citenamefont {Navr\'atil}\ \emph {et~al.}(2016)\citenamefont
  {Navr\'atil}, \citenamefont {Quaglioni}, \citenamefont {Hupin}, \citenamefont
  {Romero-Redondo},\ and\ \citenamefont {Calci}}]{NaQu16}%
  \BibitemOpen
  \bibfield  {author} {\bibinfo {author} {\bibfnamefont {P.}~\bibnamefont
  {Navr\'atil}}, \bibinfo {author} {\bibfnamefont {S.}~\bibnamefont
  {Quaglioni}}, \bibinfo {author} {\bibfnamefont {G.}~\bibnamefont {Hupin}},
  \bibinfo {author} {\bibfnamefont {C.}~\bibnamefont {Romero-Redondo}}, \ and\
  \bibinfo {author} {\bibfnamefont {A.}~\bibnamefont {Calci}},\ }\href
  {http://stacks.iop.org/1402-4896/91/i=5/a=053002} {\bibfield  {journal}
  {\bibinfo  {journal} {Phys. Scr.}\ }\textbf {\bibinfo {volume} {91}},\
  \bibinfo {pages} {053002} (\bibinfo {year} {2016})}\BibitemShut {NoStop}%
\bibitem [{\citenamefont {Entem}\ and\ \citenamefont
  {Machleidt}(2003)}]{EnMa03}%
  \BibitemOpen
  \bibfield  {author} {\bibinfo {author} {\bibfnamefont {D.~R.}\ \bibnamefont
  {Entem}}\ and\ \bibinfo {author} {\bibfnamefont {R.}~\bibnamefont
  {Machleidt}},\ }\href@noop {} {\bibfield  {journal} {\bibinfo  {journal}
  {Phys. Rev. C}\ }\textbf {\bibinfo {volume} {68}},\ \bibinfo {pages}
  {041001(R)} (\bibinfo {year} {2003})}\BibitemShut {NoStop}%
\bibitem [{\citenamefont {Navr{\'a}til}(2007)}]{Navr07}%
  \BibitemOpen
  \bibfield  {author} {\bibinfo {author} {\bibfnamefont {P.}~\bibnamefont
  {Navr{\'a}til}},\ }\href {\doibase 10.1007/s00601-007-0193-3} {\bibfield
  {journal} {\bibinfo  {journal} {Few-Body Syst.}\ }\textbf {\bibinfo {volume}
  {41}},\ \bibinfo {pages} {117} (\bibinfo {year} {2007})}\BibitemShut
  {NoStop}%
\bibitem [{\citenamefont {Gazit}\ \emph {et~al.}(2009)\citenamefont {Gazit},
  \citenamefont {Quaglioni},\ and\ \citenamefont {Navr\'atil}}]{GaQu09}%
  \BibitemOpen
  \bibfield  {author} {\bibinfo {author} {\bibfnamefont {D.}~\bibnamefont
  {Gazit}}, \bibinfo {author} {\bibfnamefont {S.}~\bibnamefont {Quaglioni}}, \
  and\ \bibinfo {author} {\bibfnamefont {P.}~\bibnamefont {Navr\'atil}},\
  }\href {\doibase 10.1103/PhysRevLett.103.102502} {\bibfield  {journal}
  {\bibinfo  {journal} {Phys. Rev. Lett.}\ }\textbf {\bibinfo {volume} {103}},\
  \bibinfo {pages} {102502} (\bibinfo {year} {2009})}\BibitemShut {NoStop}%
\bibitem [{\citenamefont {Roth}\ \emph {et~al.}(2012)\citenamefont {Roth},
  \citenamefont {Binder}, \citenamefont {Vobig}, \citenamefont {Calci},
  \citenamefont {Langhammer},\ and\ \citenamefont {Navr\'atil}}]{RoBi12}%
  \BibitemOpen
  \bibfield  {author} {\bibinfo {author} {\bibfnamefont {R.}~\bibnamefont
  {Roth}}, \bibinfo {author} {\bibfnamefont {S.}~\bibnamefont {Binder}},
  \bibinfo {author} {\bibfnamefont {K.}~\bibnamefont {Vobig}}, \bibinfo
  {author} {\bibfnamefont {A.}~\bibnamefont {Calci}}, \bibinfo {author}
  {\bibfnamefont {J.}~\bibnamefont {Langhammer}}, \ and\ \bibinfo {author}
  {\bibfnamefont {P.}~\bibnamefont {Navr\'atil}},\ }\href {\doibase
  10.1103/PhysRevLett.109.052501} {\bibfield  {journal} {\bibinfo  {journal}
  {Phys. Rev. Lett.}\ }\textbf {\bibinfo {volume} {109}},\ \bibinfo {pages}
  {052501} (\bibinfo {year} {2012})}\BibitemShut {NoStop}%
\bibitem [{\citenamefont {Roth}\ \emph {et~al.}(2014)\citenamefont {Roth},
  \citenamefont {Calci}, \citenamefont {Langhammer},\ and\ \citenamefont
  {Binder}}]{RoCa14}%
  \BibitemOpen
  \bibfield  {author} {\bibinfo {author} {\bibfnamefont {R.}~\bibnamefont
  {Roth}}, \bibinfo {author} {\bibfnamefont {A.}~\bibnamefont {Calci}},
  \bibinfo {author} {\bibfnamefont {J.}~\bibnamefont {Langhammer}}, \ and\
  \bibinfo {author} {\bibfnamefont {S.}~\bibnamefont {Binder}},\ }\href
  {\doibase 10.1103/PhysRevC.90.024325} {\bibfield  {journal} {\bibinfo
  {journal} {Phys. Rev. C}\ }\textbf {\bibinfo {volume} {90}},\ \bibinfo
  {pages} {024325} (\bibinfo {year} {2014})}\BibitemShut {NoStop}%
\bibitem [{\citenamefont {Ekstr\"om}\ \emph {et~al.}(2015)\citenamefont
  {Ekstr\"om}, \citenamefont {Jansen}, \citenamefont {Wendt}, \citenamefont
  {Hagen}, \citenamefont {Papenbrock}, \citenamefont {Carlsson}, \citenamefont
  {Forss\'en}, \citenamefont {Hjorth-Jensen}, \citenamefont {Navr\'atil},\ and\
  \citenamefont {Nazarewicz}}]{EkJa15}%
  \BibitemOpen
  \bibfield  {author} {\bibinfo {author} {\bibfnamefont {A.}~\bibnamefont
  {Ekstr\"om}}, \bibinfo {author} {\bibfnamefont {G.~R.}\ \bibnamefont
  {Jansen}}, \bibinfo {author} {\bibfnamefont {K.~A.}\ \bibnamefont {Wendt}},
  \bibinfo {author} {\bibfnamefont {G.}~\bibnamefont {Hagen}}, \bibinfo
  {author} {\bibfnamefont {T.}~\bibnamefont {Papenbrock}}, \bibinfo {author}
  {\bibfnamefont {B.~D.}\ \bibnamefont {Carlsson}}, \bibinfo {author}
  {\bibfnamefont {C.}~\bibnamefont {Forss\'en}}, \bibinfo {author}
  {\bibfnamefont {M.}~\bibnamefont {Hjorth-Jensen}}, \bibinfo {author}
  {\bibfnamefont {P.}~\bibnamefont {Navr\'atil}}, \ and\ \bibinfo {author}
  {\bibfnamefont {W.}~\bibnamefont {Nazarewicz}},\ }\href {\doibase
  10.1103/PhysRevC.91.051301} {\bibfield  {journal} {\bibinfo  {journal} {Phys.
  Rev. C}\ }\textbf {\bibinfo {volume} {91}},\ \bibinfo {pages} {051301(R)}
  (\bibinfo {year} {2015})}\BibitemShut {NoStop}%
\bibitem [{\citenamefont {Hagen}\ \emph {et~al.}(2016)\citenamefont {Hagen},
  \citenamefont {Ekstrom}, \citenamefont {Forssen}, \citenamefont {Jansen},
  \citenamefont {Nazarewicz}, \citenamefont {Papenbrock}, \citenamefont
  {Wendt}, \citenamefont {Bacca}, \citenamefont {Barnea}, \citenamefont
  {Carlsson}, \citenamefont {Drischler}, \citenamefont {Hebeler}, \citenamefont
  {Hjorth-Jensen}, \citenamefont {Miorelli}, \citenamefont {Orlandini},
  \citenamefont {Schwenk},\ and\ \citenamefont {Simonis}}]{HaEk16}%
  \BibitemOpen
  \bibfield  {author} {\bibinfo {author} {\bibfnamefont {G.}~\bibnamefont
  {Hagen}}, \bibinfo {author} {\bibfnamefont {A.}~\bibnamefont {Ekstrom}},
  \bibinfo {author} {\bibfnamefont {C.}~\bibnamefont {Forssen}}, \bibinfo
  {author} {\bibfnamefont {G.~R.}\ \bibnamefont {Jansen}}, \bibinfo {author}
  {\bibfnamefont {W.}~\bibnamefont {Nazarewicz}}, \bibinfo {author}
  {\bibfnamefont {T.}~\bibnamefont {Papenbrock}}, \bibinfo {author}
  {\bibfnamefont {K.~A.}\ \bibnamefont {Wendt}}, \bibinfo {author}
  {\bibfnamefont {S.}~\bibnamefont {Bacca}}, \bibinfo {author} {\bibfnamefont
  {N.}~\bibnamefont {Barnea}}, \bibinfo {author} {\bibfnamefont
  {B.}~\bibnamefont {Carlsson}}, \bibinfo {author} {\bibfnamefont
  {C.}~\bibnamefont {Drischler}}, \bibinfo {author} {\bibfnamefont
  {K.}~\bibnamefont {Hebeler}}, \bibinfo {author} {\bibfnamefont
  {M.}~\bibnamefont {Hjorth-Jensen}}, \bibinfo {author} {\bibfnamefont
  {M.}~\bibnamefont {Miorelli}}, \bibinfo {author} {\bibfnamefont
  {G.}~\bibnamefont {Orlandini}}, \bibinfo {author} {\bibfnamefont
  {A.}~\bibnamefont {Schwenk}}, \ and\ \bibinfo {author} {\bibfnamefont
  {J.}~\bibnamefont {Simonis}},\ }\href {http://dx.doi.org/10.1038/nphys3529}
  {\bibfield  {journal} {\bibinfo  {journal} {Nat. Phys.}\ }\textbf {\bibinfo
  {volume} {12}},\ \bibinfo {pages} {186} (\bibinfo {year} {2016})}\BibitemShut
  {NoStop}%
\bibitem [{\citenamefont {Miorelli}\ \emph {et~al.}(2016)\citenamefont
  {Miorelli}, \citenamefont {Bacca}, \citenamefont {Barnea}, \citenamefont
  {Hagen}, \citenamefont {Jansen}, \citenamefont {Orlandini},\ and\
  \citenamefont {Papenbrock}}]{MiBa16}%
  \BibitemOpen
  \bibfield  {author} {\bibinfo {author} {\bibfnamefont {M.}~\bibnamefont
  {Miorelli}}, \bibinfo {author} {\bibfnamefont {S.}~\bibnamefont {Bacca}},
  \bibinfo {author} {\bibfnamefont {N.}~\bibnamefont {Barnea}}, \bibinfo
  {author} {\bibfnamefont {G.}~\bibnamefont {Hagen}}, \bibinfo {author}
  {\bibfnamefont {G.~R.}\ \bibnamefont {Jansen}}, \bibinfo {author}
  {\bibfnamefont {G.}~\bibnamefont {Orlandini}}, \ and\ \bibinfo {author}
  {\bibfnamefont {T.}~\bibnamefont {Papenbrock}},\ }\href {\doibase
  10.1103/PhysRevC.94.034317} {\bibfield  {journal} {\bibinfo  {journal} {Phys.
  Rev. C}\ }\textbf {\bibinfo {volume} {94}},\ \bibinfo {pages} {034317}
  (\bibinfo {year} {2016})}\BibitemShut {NoStop}%
\bibitem [{\citenamefont {Descouvemont}\ and\ \citenamefont
  {Baye}(2010)}]{DeBa10}%
  \BibitemOpen
  \bibfield  {author} {\bibinfo {author} {\bibfnamefont {P.}~\bibnamefont
  {Descouvemont}}\ and\ \bibinfo {author} {\bibfnamefont {D.}~\bibnamefont
  {Baye}},\ }\href {http://stacks.iop.org/0034-4885/73/i=3/a=036301} {\bibfield
   {journal} {\bibinfo  {journal} {Rep. Prog. in Phys.}\ }\textbf {\bibinfo
  {volume} {73}},\ \bibinfo {pages} {036301} (\bibinfo {year}
  {2010})}\BibitemShut {NoStop}%
\bibitem [{\citenamefont {Hesse}\ \emph {et~al.}(2002)\citenamefont {Hesse},
  \citenamefont {Roland},\ and\ \citenamefont {Baye}}]{HeRo02}%
  \BibitemOpen
  \bibfield  {author} {\bibinfo {author} {\bibfnamefont {M.}~\bibnamefont
  {Hesse}}, \bibinfo {author} {\bibfnamefont {J.}~\bibnamefont {Roland}}, \
  and\ \bibinfo {author} {\bibfnamefont {D.}~\bibnamefont {Baye}},\ }\href
  {\doibase http://dx.doi.org/10.1016/S0375-9474(02)01040-0} {\bibfield
  {journal} {\bibinfo  {journal} {Nucl. Phys. A}\ }\textbf {\bibinfo {volume}
  {709}},\ \bibinfo {pages} {184 } (\bibinfo {year} {2002})}\BibitemShut
  {NoStop}%
\bibitem [{\citenamefont {Baye}\ \emph {et~al.}(2002)\citenamefont {Baye},
  \citenamefont {Goldbeter},\ and\ \citenamefont {Sparenberg}}]{BaGo02}%
  \BibitemOpen
  \bibfield  {author} {\bibinfo {author} {\bibfnamefont {D.}~\bibnamefont
  {Baye}}, \bibinfo {author} {\bibfnamefont {J.}~\bibnamefont {Goldbeter}}, \
  and\ \bibinfo {author} {\bibfnamefont {J.-M.}\ \bibnamefont {Sparenberg}},\
  }\href {\doibase 10.1103/PhysRevA.65.052710} {\bibfield  {journal} {\bibinfo
  {journal} {Phys. Rev. A}\ }\textbf {\bibinfo {volume} {65}},\ \bibinfo
  {pages} {052710} (\bibinfo {year} {2002})}\BibitemShut {NoStop}%
\bibitem [{\citenamefont {Schneider}(1981)}]{Schn81}%
  \BibitemOpen
  \bibfield  {author} {\bibinfo {author} {\bibfnamefont {B.~I.}\ \bibnamefont
  {Schneider}},\ }\href {\doibase 10.1103/PhysRevA.24.1} {\bibfield  {journal}
  {\bibinfo  {journal} {Phys. Rev. A}\ }\textbf {\bibinfo {volume} {24}},\
  \bibinfo {pages} {1} (\bibinfo {year} {1981})}\BibitemShut {NoStop}%
\bibitem [{\citenamefont {Dohet-Eraly}\ \emph {et~al.}(2016)\citenamefont
  {Dohet-Eraly}, \citenamefont {Navr\'atil}, \citenamefont {Quaglioni},
  \citenamefont {Horiuchi}, \citenamefont {Hupin},\ and\ \citenamefont
  {Raimondi}}]{DoNa16}%
  \BibitemOpen
  \bibfield  {author} {\bibinfo {author} {\bibfnamefont {J.}~\bibnamefont
  {Dohet-Eraly}}, \bibinfo {author} {\bibfnamefont {P.}~\bibnamefont
  {Navr\'atil}}, \bibinfo {author} {\bibfnamefont {S.}~\bibnamefont
  {Quaglioni}}, \bibinfo {author} {\bibfnamefont {W.}~\bibnamefont {Horiuchi}},
  \bibinfo {author} {\bibfnamefont {G.}~\bibnamefont {Hupin}}, \ and\ \bibinfo
  {author} {\bibfnamefont {F.}~\bibnamefont {Raimondi}},\ }\href {\doibase
  http://dx.doi.org/10.1016/j.physletb.2016.04.021} {\bibfield  {journal}
  {\bibinfo  {journal} {Phys. Lett. B}\ }\textbf {\bibinfo {volume} {757}},\
  \bibinfo {pages} {430 } (\bibinfo {year} {2016})}\BibitemShut {NoStop}%
\bibitem [{\citenamefont {Hupin}\ \emph {et~al.}(2013)\citenamefont {Hupin},
  \citenamefont {Langhammer}, \citenamefont {Navr\'atil}, \citenamefont
  {Quaglioni}, \citenamefont {Calci},\ and\ \citenamefont {Roth}}]{HuLa13}%
  \BibitemOpen
  \bibfield  {author} {\bibinfo {author} {\bibfnamefont {G.}~\bibnamefont
  {Hupin}}, \bibinfo {author} {\bibfnamefont {J.}~\bibnamefont {Langhammer}},
  \bibinfo {author} {\bibfnamefont {P.}~\bibnamefont {Navr\'atil}}, \bibinfo
  {author} {\bibfnamefont {S.}~\bibnamefont {Quaglioni}}, \bibinfo {author}
  {\bibfnamefont {A.}~\bibnamefont {Calci}}, \ and\ \bibinfo {author}
  {\bibfnamefont {R.}~\bibnamefont {Roth}},\ }\href {\doibase
  10.1103/PhysRevC.88.054622} {\bibfield  {journal} {\bibinfo  {journal} {Phys.
  Rev. C}\ }\textbf {\bibinfo {volume} {88}},\ \bibinfo {pages} {054622}
  (\bibinfo {year} {2013})}\BibitemShut {NoStop}%
\bibitem [{\citenamefont {Roth}(2009)}]{Roth09}%
  \BibitemOpen
  \bibfield  {author} {\bibinfo {author} {\bibfnamefont {R.}~\bibnamefont
  {Roth}},\ }\href {\doibase 10.1103/PhysRevC.79.064324} {\bibfield  {journal}
  {\bibinfo  {journal} {Phys. Rev. C}\ }\textbf {\bibinfo {volume} {79}},\
  \bibinfo {pages} {064324} (\bibinfo {year} {2009})}\BibitemShut {NoStop}%
\bibitem [{\citenamefont {Roth}\ and\ \citenamefont
  {Navr\'atil}(2007)}]{RoNa07}%
  \BibitemOpen
  \bibfield  {author} {\bibinfo {author} {\bibfnamefont {R.}~\bibnamefont
  {Roth}}\ and\ \bibinfo {author} {\bibfnamefont {P.}~\bibnamefont
  {Navr\'atil}},\ }\href@noop {} {\bibfield  {journal} {\bibinfo  {journal}
  {Phys. Rev. Lett.}\ }\textbf {\bibinfo {volume} {99}},\ \bibinfo {pages}
  {092501} (\bibinfo {year} {2007})}\BibitemShut {NoStop}%
\bibitem [{\citenamefont {Langhammer}\ \emph {et~al.}(2015)\citenamefont
  {Langhammer}, \citenamefont {Navr\'atil}, \citenamefont {Quaglioni},
  \citenamefont {Hupin}, \citenamefont {Calci},\ and\ \citenamefont
  {Roth}}]{LaNa15}%
  \BibitemOpen
  \bibfield  {author} {\bibinfo {author} {\bibfnamefont {J.}~\bibnamefont
  {Langhammer}}, \bibinfo {author} {\bibfnamefont {P.}~\bibnamefont
  {Navr\'atil}}, \bibinfo {author} {\bibfnamefont {S.}~\bibnamefont
  {Quaglioni}}, \bibinfo {author} {\bibfnamefont {G.}~\bibnamefont {Hupin}},
  \bibinfo {author} {\bibfnamefont {A.}~\bibnamefont {Calci}}, \ and\ \bibinfo
  {author} {\bibfnamefont {R.}~\bibnamefont {Roth}},\ }\href {\doibase
  10.1103/PhysRevC.91.021301} {\bibfield  {journal} {\bibinfo  {journal} {Phys.
  Rev. C}\ }\textbf {\bibinfo {volume} {91}},\ \bibinfo {pages} {021301(R)}
  (\bibinfo {year} {2015})}\BibitemShut {NoStop}%
\bibitem [{\citenamefont {Binder}\ \emph {et~al.}(2014)\citenamefont {Binder},
  \citenamefont {Langhammer}, \citenamefont {Calci},\ and\ \citenamefont
  {Roth}}]{BiLa13b}%
  \BibitemOpen
  \bibfield  {author} {\bibinfo {author} {\bibfnamefont {S.}~\bibnamefont
  {Binder}}, \bibinfo {author} {\bibfnamefont {J.}~\bibnamefont {Langhammer}},
  \bibinfo {author} {\bibfnamefont {A.}~\bibnamefont {Calci}}, \ and\ \bibinfo
  {author} {\bibfnamefont {R.}~\bibnamefont {Roth}},\ }\href {\doibase
  http://dx.doi.org/10.1016/j.physletb.2014.07.010} {\bibfield  {journal}
  {\bibinfo  {journal} {Phys. Lett. B}\ }\textbf {\bibinfo {volume} {736}},\
  \bibinfo {pages} {119 } (\bibinfo {year} {2014})}\BibitemShut {NoStop}%
\bibitem [{\citenamefont {Roth}\ \emph {et~al.}(2011)\citenamefont {Roth},
  \citenamefont {Langhammer}, \citenamefont {Calci}, \citenamefont {Binder},\
  and\ \citenamefont {Navr{\'a}til}}]{RoLa11}%
  \BibitemOpen
  \bibfield  {author} {\bibinfo {author} {\bibfnamefont {R.}~\bibnamefont
  {Roth}}, \bibinfo {author} {\bibfnamefont {J.}~\bibnamefont {Langhammer}},
  \bibinfo {author} {\bibfnamefont {A.}~\bibnamefont {Calci}}, \bibinfo
  {author} {\bibfnamefont {S.}~\bibnamefont {Binder}}, \ and\ \bibinfo {author}
  {\bibfnamefont {P.}~\bibnamefont {Navr{\'a}til}},\ }\href {\doibase
  10.1103/PhysRevLett.107.072501} {\bibfield  {journal} {\bibinfo  {journal}
  {Phys. Rev. Lett.}\ }\textbf {\bibinfo {volume} {107}},\ \bibinfo {pages}
  {072501} (\bibinfo {year} {2011})}\BibitemShut {NoStop}%
\bibitem [{\citenamefont {Wegner}(1994)}]{Wegn94}%
  \BibitemOpen
  \bibfield  {author} {\bibinfo {author} {\bibfnamefont {F.}~\bibnamefont
  {Wegner}},\ }\href@noop {} {\bibfield  {journal} {\bibinfo  {journal} {Ann.
  Phys. (Berlin)}\ }\textbf {\bibinfo {volume} {506}},\ \bibinfo {pages} {77}
  (\bibinfo {year} {1994})}\BibitemShut {NoStop}%
\bibitem [{\citenamefont {Bogner}\ \emph {et~al.}(2007)\citenamefont {Bogner},
  \citenamefont {Furnstahl},\ and\ \citenamefont {Perry}}]{BoFu07}%
  \BibitemOpen
  \bibfield  {author} {\bibinfo {author} {\bibfnamefont {S.~K.}\ \bibnamefont
  {Bogner}}, \bibinfo {author} {\bibfnamefont {R.~J.}\ \bibnamefont
  {Furnstahl}}, \ and\ \bibinfo {author} {\bibfnamefont {R.~J.}\ \bibnamefont
  {Perry}},\ }\href@noop {} {\bibfield  {journal} {\bibinfo  {journal} {Phys.
  Rev. C}\ }\textbf {\bibinfo {volume} {75}},\ \bibinfo {pages} {061001(R)}
  (\bibinfo {year} {2007})}\BibitemShut {NoStop}%
\bibitem [{\citenamefont {Szpigel}\ and\ \citenamefont {Perry}(2000)}]{SzPe00}%
  \BibitemOpen
  \bibfield  {author} {\bibinfo {author} {\bibfnamefont {S.}~\bibnamefont
  {Szpigel}}\ and\ \bibinfo {author} {\bibfnamefont {R.~J.}\ \bibnamefont
  {Perry}},\ }in\ \href@noop {} {\emph {\bibinfo {booktitle} {Quantum {F}ield
  {T}heory. {A} 20th {C}entury {P}rofile}}},\ \bibinfo {editor} {edited by\
  \bibinfo {editor} {\bibnamefont {{A. N. Mitra}}}}\ (\bibinfo  {publisher}
  {Hindustan Publishing Co., New Delhi},\ \bibinfo {year} {2000})\BibitemShut
  {NoStop}%
\bibitem [{Sup()}]{Supplement}%
  \BibitemOpen
  \href@noop {} {}\bibinfo {note} {See Supplemental Material for
  details of the calculations, which includes
  Refs.~\cite{HeBi13,HeBo13,CiBa13,GeCa16}}\BibitemShut {NoStop}%
\bibitem [{\citenamefont {Aoi}\ \emph {et~al.}(1997)\citenamefont {Aoi},
  \citenamefont {Yoneda}, \citenamefont {Miyatake}, \citenamefont {Ogawa},
  \citenamefont {Yamamoto}, \citenamefont {Ideguchi}, \citenamefont {Kishida},
  \citenamefont {Nakamura}, \citenamefont {Notani}, \citenamefont {Sakurai},
  \citenamefont {Teranishi}, \citenamefont {Wu}, \citenamefont {Yamamoto},
  \citenamefont {Watanabe}, \citenamefont {Yoshida},\ and\ \citenamefont
  {Ishihara}}]{AoYo97}%
  \BibitemOpen
  \bibfield  {author} {\bibinfo {author} {\bibfnamefont {N.}~\bibnamefont
  {Aoi}}, \bibinfo {author} {\bibfnamefont {K.}~\bibnamefont {Yoneda}},
  \bibinfo {author} {\bibfnamefont {H.}~\bibnamefont {Miyatake}}, \bibinfo
  {author} {\bibfnamefont {H.}~\bibnamefont {Ogawa}}, \bibinfo {author}
  {\bibfnamefont {Y.}~\bibnamefont {Yamamoto}}, \bibinfo {author}
  {\bibfnamefont {E.}~\bibnamefont {Ideguchi}}, \bibinfo {author}
  {\bibfnamefont {T.}~\bibnamefont {Kishida}}, \bibinfo {author} {\bibfnamefont
  {T.}~\bibnamefont {Nakamura}}, \bibinfo {author} {\bibfnamefont
  {M.}~\bibnamefont {Notani}}, \bibinfo {author} {\bibfnamefont
  {H.}~\bibnamefont {Sakurai}}, \bibinfo {author} {\bibfnamefont
  {T.}~\bibnamefont {Teranishi}}, \bibinfo {author} {\bibfnamefont
  {H.}~\bibnamefont {Wu}}, \bibinfo {author} {\bibfnamefont {S.}~\bibnamefont
  {Yamamoto}}, \bibinfo {author} {\bibfnamefont {Y.}~\bibnamefont {Watanabe}},
  \bibinfo {author} {\bibfnamefont {A.}~\bibnamefont {Yoshida}}, \ and\
  \bibinfo {author} {\bibfnamefont {M.}~\bibnamefont {Ishihara}},\ }\href
  {\doibase http://dx.doi.org/10.1016/S0375-9474(97)00087-0} {\bibfield
  {journal} {\bibinfo  {journal} {Nucl. Phys. A}\ }\textbf {\bibinfo {volume}
  {616}},\ \bibinfo {pages} {181 } (\bibinfo {year} {1997})}\BibitemShut
  {NoStop}%
\bibitem [{\citenamefont {Hirayama}\ \emph {et~al.}(2004)\citenamefont
  {Hirayama}, \citenamefont {Shimoda}, \citenamefont {Izumi}, \citenamefont
  {Yano}, \citenamefont {Yagi}, \citenamefont {Hatakeyama}, \citenamefont
  {Levy}, \citenamefont {Jackson},\ and\ \citenamefont {Miyatake}}]{HiSh04}%
  \BibitemOpen
  \bibfield  {author} {\bibinfo {author} {\bibfnamefont {Y.}~\bibnamefont
  {Hirayama}}, \bibinfo {author} {\bibfnamefont {T.}~\bibnamefont {Shimoda}},
  \bibinfo {author} {\bibfnamefont {H.}~\bibnamefont {Izumi}}, \bibinfo
  {author} {\bibfnamefont {H.}~\bibnamefont {Yano}}, \bibinfo {author}
  {\bibfnamefont {M.}~\bibnamefont {Yagi}}, \bibinfo {author} {\bibfnamefont
  {A.}~\bibnamefont {Hatakeyama}}, \bibinfo {author} {\bibfnamefont
  {C.}~\bibnamefont {Levy}}, \bibinfo {author} {\bibfnamefont {K.}~\bibnamefont
  {Jackson}}, \ and\ \bibinfo {author} {\bibfnamefont {H.}~\bibnamefont
  {Miyatake}},\ }\href {\doibase
  http://dx.doi.org/10.1016/j.nuclphysa.2004.04.032} {\bibfield  {journal}
  {\bibinfo  {journal} {Nucl. Phys. A}\ }\textbf {\bibinfo {volume} {738}},\
  \bibinfo {pages} {201 } (\bibinfo {year} {2004})}\BibitemShut {NoStop}%
\bibitem [{\citenamefont {Fossez}\ \emph {et~al.}(2016)\citenamefont {Fossez},
  \citenamefont {Nazarewicz}, \citenamefont {Jaganathen}, \citenamefont
  {Michel},\ and\ \citenamefont {P\l{}oszajczak}}]{FoNa16b}%
  \BibitemOpen
  \bibfield  {author} {\bibinfo {author} {\bibfnamefont {K.}~\bibnamefont
  {Fossez}}, \bibinfo {author} {\bibfnamefont {W.}~\bibnamefont {Nazarewicz}},
  \bibinfo {author} {\bibfnamefont {Y.}~\bibnamefont {Jaganathen}}, \bibinfo
  {author} {\bibfnamefont {N.}~\bibnamefont {Michel}}, \ and\ \bibinfo {author}
  {\bibfnamefont {M.}~\bibnamefont {P\l{}oszajczak}},\ }\href {\doibase
  10.1103/PhysRevC.93.011305} {\bibfield  {journal} {\bibinfo  {journal} {Phys.
  Rev. C}\ }\textbf {\bibinfo {volume} {93}},\ \bibinfo {pages} {011305}
  (\bibinfo {year} {2016})}\BibitemShut {NoStop}%
\bibitem [{\citenamefont {Raimondi}\ \emph {et~al.}(2016)\citenamefont
  {Raimondi}, \citenamefont {Hupin}, \citenamefont {Navr\'atil},\ and\
  \citenamefont {Quaglioni}}]{RaHu16}%
  \BibitemOpen
  \bibfield  {author} {\bibinfo {author} {\bibfnamefont {F.}~\bibnamefont
  {Raimondi}}, \bibinfo {author} {\bibfnamefont {G.}~\bibnamefont {Hupin}},
  \bibinfo {author} {\bibfnamefont {P.}~\bibnamefont {Navr\'atil}}, \ and\
  \bibinfo {author} {\bibfnamefont {S.}~\bibnamefont {Quaglioni}},\ }\href
  {\doibase 10.1103/PhysRevC.93.054606} {\bibfield  {journal} {\bibinfo
  {journal} {Phys. Rev. C}\ }\textbf {\bibinfo {volume} {93}},\ \bibinfo
  {pages} {054606} (\bibinfo {year} {2016})}\BibitemShut {NoStop}%
\bibitem [{\citenamefont {Schuster}\ \emph {et~al.}(2014)\citenamefont
  {Schuster}, \citenamefont {Quaglioni}, \citenamefont {Johnson}, \citenamefont
  {Jurgenson},\ and\ \citenamefont {Navr\'atil}}]{ScQu14}%
  \BibitemOpen
  \bibfield  {author} {\bibinfo {author} {\bibfnamefont {M.~D.}\ \bibnamefont
  {Schuster}}, \bibinfo {author} {\bibfnamefont {S.}~\bibnamefont {Quaglioni}},
  \bibinfo {author} {\bibfnamefont {C.~W.}\ \bibnamefont {Johnson}}, \bibinfo
  {author} {\bibfnamefont {E.~D.}\ \bibnamefont {Jurgenson}}, \ and\ \bibinfo
  {author} {\bibfnamefont {P.}~\bibnamefont {Navr\'atil}},\ }\href {\doibase
  10.1103/PhysRevC.90.011301} {\bibfield  {journal} {\bibinfo  {journal} {Phys.
  Rev. C}\ }\textbf {\bibinfo {volume} {90}},\ \bibinfo {pages} {011301(R)}
  (\bibinfo {year} {2014})}\BibitemShut {NoStop}%
\bibitem [{\citenamefont {Schuster}\ \emph {et~al.}(2015)\citenamefont
  {Schuster}, \citenamefont {Quaglioni}, \citenamefont {Johnson}, \citenamefont
  {Jurgenson},\ and\ \citenamefont {Navr\'atil}}]{ScQu15}%
  \BibitemOpen
  \bibfield  {author} {\bibinfo {author} {\bibfnamefont {M.~D.}\ \bibnamefont
  {Schuster}}, \bibinfo {author} {\bibfnamefont {S.}~\bibnamefont {Quaglioni}},
  \bibinfo {author} {\bibfnamefont {C.~W.}\ \bibnamefont {Johnson}}, \bibinfo
  {author} {\bibfnamefont {E.~D.}\ \bibnamefont {Jurgenson}}, \ and\ \bibinfo
  {author} {\bibfnamefont {P.}~\bibnamefont {Navr\'atil}},\ }\href {\doibase
  10.1103/PhysRevC.92.014320} {\bibfield  {journal} {\bibinfo  {journal} {Phys.
  Rev. C}\ }\textbf {\bibinfo {volume} {92}},\ \bibinfo {pages} {014320}
  (\bibinfo {year} {2015})}\BibitemShut {NoStop}%
\bibitem [{\citenamefont {Palit}\ \emph {et~al.}(2003)\citenamefont {Palit},
  \citenamefont {Adrich}, \citenamefont {Aumann}, \citenamefont {Boretzky},
  \citenamefont {Carlson}, \citenamefont {Cortina}, \citenamefont
  {Datta~Pramanik}, \citenamefont {Elze}, \citenamefont {Emling}, \citenamefont
  {Geissel}, \citenamefont {Hellstr\"om}, \citenamefont {Jones}, \citenamefont
  {Kratz}, \citenamefont {Kulessa}, \citenamefont {Leifels}, \citenamefont
  {Leistenschneider}, \citenamefont {M\"unzenberg}, \citenamefont {Nociforo},
  \citenamefont {Reiter}, \citenamefont {Simon}, \citenamefont {S\"ummerer},\
  and\ \citenamefont {Walus}}]{PaAd03}%
  \BibitemOpen
  \bibfield  {author} {\bibinfo {author} {\bibfnamefont {R.}~\bibnamefont
  {Palit}}, \bibinfo {author} {\bibfnamefont {P.}~\bibnamefont {Adrich}},
  \bibinfo {author} {\bibfnamefont {T.}~\bibnamefont {Aumann}}, \bibinfo
  {author} {\bibfnamefont {K.}~\bibnamefont {Boretzky}}, \bibinfo {author}
  {\bibfnamefont {B.~V.}\ \bibnamefont {Carlson}}, \bibinfo {author}
  {\bibfnamefont {D.}~\bibnamefont {Cortina}}, \bibinfo {author} {\bibfnamefont
  {U.}~\bibnamefont {Datta~Pramanik}}, \bibinfo {author} {\bibfnamefont
  {T.~W.}\ \bibnamefont {Elze}}, \bibinfo {author} {\bibfnamefont
  {H.}~\bibnamefont {Emling}}, \bibinfo {author} {\bibfnamefont
  {H.}~\bibnamefont {Geissel}}, \bibinfo {author} {\bibfnamefont
  {M.}~\bibnamefont {Hellstr\"om}}, \bibinfo {author} {\bibfnamefont {K.~L.}\
  \bibnamefont {Jones}}, \bibinfo {author} {\bibfnamefont {J.~V.}\ \bibnamefont
  {Kratz}}, \bibinfo {author} {\bibfnamefont {R.}~\bibnamefont {Kulessa}},
  \bibinfo {author} {\bibfnamefont {Y.}~\bibnamefont {Leifels}}, \bibinfo
  {author} {\bibfnamefont {A.}~\bibnamefont {Leistenschneider}}, \bibinfo
  {author} {\bibfnamefont {G.}~\bibnamefont {M\"unzenberg}}, \bibinfo {author}
  {\bibfnamefont {C.}~\bibnamefont {Nociforo}}, \bibinfo {author}
  {\bibfnamefont {P.}~\bibnamefont {Reiter}}, \bibinfo {author} {\bibfnamefont
  {H.}~\bibnamefont {Simon}}, \bibinfo {author} {\bibfnamefont
  {K.}~\bibnamefont {S\"ummerer}}, \ and\ \bibinfo {author} {\bibfnamefont
  {W.}~\bibnamefont {Walus}} (\bibinfo {collaboration} {LAND/FRS
  Collaboration}),\ }\href {\doibase 10.1103/PhysRevC.68.034318} {\bibfield
  {journal} {\bibinfo  {journal} {Phys. Rev. C}\ }\textbf {\bibinfo {volume}
  {68}},\ \bibinfo {pages} {034318} (\bibinfo {year} {2003})}\BibitemShut
  {NoStop}%
\bibitem [{\citenamefont {Aumann}\ and\ \citenamefont
  {Nakamura}(2013)}]{AuNa13}%
  \BibitemOpen
  \bibfield  {author} {\bibinfo {author} {\bibfnamefont {T.}~\bibnamefont
  {Aumann}}\ and\ \bibinfo {author} {\bibfnamefont {T.}~\bibnamefont
  {Nakamura}},\ }\href {http://stacks.iop.org/1402-4896/2013/i=T152/a=014012}
  {\bibfield  {journal} {\bibinfo  {journal} {Phys. Scr.}\ }\textbf {\bibinfo
  {volume} {T152}},\ \bibinfo {pages} {014012} (\bibinfo {year}
  {2013})}\BibitemShut {NoStop}%
\bibitem [{\citenamefont {{Nakamura}}\ \emph {et~al.}(1994)\citenamefont
  {{Nakamura}}, \citenamefont {{Shimoura}}, \citenamefont {{Kobayashi}},
  \citenamefont {{Teranishi}}, \citenamefont {{Abe}}, \citenamefont {{Aoi}},
  \citenamefont {{Doki}}, \citenamefont {{Fujimaki}}, \citenamefont {{Inabe}},
  \citenamefont {{Iwasa}}, \citenamefont {{Katori}}, \citenamefont {{Kubo}},
  \citenamefont {{Okuno}}, \citenamefont {{Suzuki}}, \citenamefont
  {{Tanihata}}, \citenamefont {{Watanabe}}, \citenamefont {{Yoshida}},\ and\
  \citenamefont {{Ishihara}}}]{NaSh94}%
  \BibitemOpen
  \bibfield  {author} {\bibinfo {author} {\bibfnamefont {T.}~\bibnamefont
  {{Nakamura}}}, \bibinfo {author} {\bibfnamefont {S.}~\bibnamefont
  {{Shimoura}}}, \bibinfo {author} {\bibfnamefont {T.}~\bibnamefont
  {{Kobayashi}}}, \bibinfo {author} {\bibfnamefont {T.}~\bibnamefont
  {{Teranishi}}}, \bibinfo {author} {\bibfnamefont {K.}~\bibnamefont {{Abe}}},
  \bibinfo {author} {\bibfnamefont {N.}~\bibnamefont {{Aoi}}}, \bibinfo
  {author} {\bibfnamefont {Y.}~\bibnamefont {{Doki}}}, \bibinfo {author}
  {\bibfnamefont {M.}~\bibnamefont {{Fujimaki}}}, \bibinfo {author}
  {\bibfnamefont {N.}~\bibnamefont {{Inabe}}}, \bibinfo {author} {\bibfnamefont
  {N.}~\bibnamefont {{Iwasa}}}, \bibinfo {author} {\bibfnamefont
  {K.}~\bibnamefont {{Katori}}}, \bibinfo {author} {\bibfnamefont
  {T.}~\bibnamefont {{Kubo}}}, \bibinfo {author} {\bibfnamefont
  {H.}~\bibnamefont {{Okuno}}}, \bibinfo {author} {\bibfnamefont
  {T.}~\bibnamefont {{Suzuki}}}, \bibinfo {author} {\bibfnamefont
  {I.}~\bibnamefont {{Tanihata}}}, \bibinfo {author} {\bibfnamefont
  {Y.}~\bibnamefont {{Watanabe}}}, \bibinfo {author} {\bibfnamefont
  {A.}~\bibnamefont {{Yoshida}}}, \ and\ \bibinfo {author} {\bibfnamefont
  {M.}~\bibnamefont {{Ishihara}}},\ }\href {\doibase
  10.1016/0370-2693(94)91055-3} {\bibfield  {journal} {\bibinfo  {journal}
  {Phys. Lett. B}\ }\textbf {\bibinfo {volume} {331}},\ \bibinfo {pages} {296}
  (\bibinfo {year} {1994})}\BibitemShut {NoStop}%
\bibitem [{\citenamefont {Fukuda}\ \emph {et~al.}(2004)\citenamefont {Fukuda},
  \citenamefont {Nakamura}, \citenamefont {Aoi}, \citenamefont {Imai},
  \citenamefont {Ishihara}, \citenamefont {Kobayashi}, \citenamefont {Iwasaki},
  \citenamefont {Kubo}, \citenamefont {Mengoni}, \citenamefont {Notani},
  \citenamefont {Otsu}, \citenamefont {Sakurai}, \citenamefont {Shimoura},
  \citenamefont {Teranishi}, \citenamefont {Watanabe},\ and\ \citenamefont
  {Yoneda}}]{FuNa04}%
  \BibitemOpen
  \bibfield  {author} {\bibinfo {author} {\bibfnamefont {N.}~\bibnamefont
  {Fukuda}}, \bibinfo {author} {\bibfnamefont {T.}~\bibnamefont {Nakamura}},
  \bibinfo {author} {\bibfnamefont {N.}~\bibnamefont {Aoi}}, \bibinfo {author}
  {\bibfnamefont {N.}~\bibnamefont {Imai}}, \bibinfo {author} {\bibfnamefont
  {M.}~\bibnamefont {Ishihara}}, \bibinfo {author} {\bibfnamefont
  {T.}~\bibnamefont {Kobayashi}}, \bibinfo {author} {\bibfnamefont
  {H.}~\bibnamefont {Iwasaki}}, \bibinfo {author} {\bibfnamefont
  {T.}~\bibnamefont {Kubo}}, \bibinfo {author} {\bibfnamefont {A.}~\bibnamefont
  {Mengoni}}, \bibinfo {author} {\bibfnamefont {M.}~\bibnamefont {Notani}},
  \bibinfo {author} {\bibfnamefont {H.}~\bibnamefont {Otsu}}, \bibinfo {author}
  {\bibfnamefont {H.}~\bibnamefont {Sakurai}}, \bibinfo {author} {\bibfnamefont
  {S.}~\bibnamefont {Shimoura}}, \bibinfo {author} {\bibfnamefont
  {T.}~\bibnamefont {Teranishi}}, \bibinfo {author} {\bibfnamefont {Y.~X.}\
  \bibnamefont {Watanabe}}, \ and\ \bibinfo {author} {\bibfnamefont
  {K.}~\bibnamefont {Yoneda}},\ }\href {\doibase 10.1103/PhysRevC.70.054606}
  {\bibfield  {journal} {\bibinfo  {journal} {Phys. Rev. C}\ }\textbf {\bibinfo
  {volume} {70}},\ \bibinfo {pages} {054606} (\bibinfo {year}
  {2004})}\BibitemShut {NoStop}%
\bibitem [{\citenamefont {Descouvemont}(1997)}]{Desc97}%
  \BibitemOpen
  \bibfield  {author} {\bibinfo {author} {\bibfnamefont {P.}~\bibnamefont
  {Descouvemont}},\ }\href {\doibase
  http://dx.doi.org/10.1016/S0375-9474(97)00015-8} {\bibfield  {journal}
  {\bibinfo  {journal} {Nucl. Phys. A}\ }\textbf {\bibinfo {volume} {615}},\
  \bibinfo {pages} {261 } (\bibinfo {year} {1997})}\BibitemShut {NoStop}%
\bibitem [{IRI()}]{IRIS}%
  \BibitemOpen
  \href@noop {} {}\bibinfo {note} {A. Kumar, R. Kanungo, A. Calci, P.
  Navr{\'a}til et al. in preparation.}\BibitemShut {Stop}%
\bibitem [{\citenamefont {Hergert}\ \emph
  {et~al.}(2013{\natexlab{a}})\citenamefont {Hergert}, \citenamefont {Binder},
  \citenamefont {Calci}, \citenamefont {Langhammer},\ and\ \citenamefont
  {Roth}}]{HeBi13}%
  \BibitemOpen
  \bibfield  {author} {\bibinfo {author} {\bibfnamefont {H.}~\bibnamefont
  {Hergert}}, \bibinfo {author} {\bibfnamefont {S.}~\bibnamefont {Binder}},
  \bibinfo {author} {\bibfnamefont {A.}~\bibnamefont {Calci}}, \bibinfo
  {author} {\bibfnamefont {J.}~\bibnamefont {Langhammer}}, \ and\ \bibinfo
  {author} {\bibfnamefont {R.}~\bibnamefont {Roth}},\ }\href {\doibase
  10.1103/PhysRevLett.110.242501} {\bibfield  {journal} {\bibinfo  {journal}
  {Phys. Rev. Lett.}\ }\textbf {\bibinfo {volume} {110}},\ \bibinfo {pages}
  {242501} (\bibinfo {year} {2013}{\natexlab{a}})}\BibitemShut {NoStop}%
\bibitem [{\citenamefont {Hergert}\ \emph
  {et~al.}(2013{\natexlab{b}})\citenamefont {Hergert}, \citenamefont {Bogner},
  \citenamefont {Binder}, \citenamefont {Calci}, \citenamefont {Langhammer},
  \citenamefont {Roth},\ and\ \citenamefont {Schwenk}}]{HeBo13}%
  \BibitemOpen
  \bibfield  {author} {\bibinfo {author} {\bibfnamefont {H.}~\bibnamefont
  {Hergert}}, \bibinfo {author} {\bibfnamefont {S.~K.}\ \bibnamefont {Bogner}},
  \bibinfo {author} {\bibfnamefont {S.}~\bibnamefont {Binder}}, \bibinfo
  {author} {\bibfnamefont {A.}~\bibnamefont {Calci}}, \bibinfo {author}
  {\bibfnamefont {J.}~\bibnamefont {Langhammer}}, \bibinfo {author}
  {\bibfnamefont {R.}~\bibnamefont {Roth}}, \ and\ \bibinfo {author}
  {\bibfnamefont {A.}~\bibnamefont {Schwenk}},\ }\href {\doibase
  10.1103/PhysRevC.87.034307} {\bibfield  {journal} {\bibinfo  {journal} {Phys.
  Rev. C}\ }\textbf {\bibinfo {volume} {87}},\ \bibinfo {pages} {034307}
  (\bibinfo {year} {2013}{\natexlab{b}})}\BibitemShut {NoStop}%
\bibitem [{\citenamefont {Cipollone}\ \emph {et~al.}(2013)\citenamefont
  {Cipollone}, \citenamefont {Barbieri},\ and\ \citenamefont
  {Navr\'atil}}]{CiBa13}%
  \BibitemOpen
  \bibfield  {author} {\bibinfo {author} {\bibfnamefont {A.}~\bibnamefont
  {Cipollone}}, \bibinfo {author} {\bibfnamefont {C.}~\bibnamefont {Barbieri}},
  \ and\ \bibinfo {author} {\bibfnamefont {P.}~\bibnamefont {Navr\'atil}},\
  }\href {\doibase 10.1103/PhysRevLett.111.062501} {\bibfield  {journal}
  {\bibinfo  {journal} {Phys. Rev. Lett.}\ }\textbf {\bibinfo {volume} {111}},\
  \bibinfo {pages} {062501} (\bibinfo {year} {2013})}\BibitemShut {NoStop}%
\bibitem [{\citenamefont {Gebrerufael}\ \emph {et~al.}(2016)\citenamefont
  {Gebrerufael}, \citenamefont {Calci},\ and\ \citenamefont {Roth}}]{GeCa16}%
  \BibitemOpen
  \bibfield  {author} {\bibinfo {author} {\bibfnamefont {E.}~\bibnamefont
  {Gebrerufael}}, \bibinfo {author} {\bibfnamefont {A.}~\bibnamefont {Calci}},
  \ and\ \bibinfo {author} {\bibfnamefont {R.}~\bibnamefont {Roth}},\ }\href
  {\doibase 10.1103/PhysRevC.93.031301} {\bibfield  {journal} {\bibinfo
  {journal} {Phys. Rev. C}\ }\textbf {\bibinfo {volume} {93}},\ \bibinfo
  {pages} {031301} (\bibinfo {year} {2016})}\BibitemShut {NoStop}%
\end{thebibliography}

%


\clearpage

\section{Supplemental Material}
\subsection{Details of the calculations.}
In this work we investigate the most-commonly used chiral NN interaction obtained at next-to-next-to-next-to leading order (N$^3$LO)~\cite{EnMa03} augmented with a local 3N force at next-to-next-to leading order N$^2$LO that was initially introduced in Ref.~\cite{Navr07} and constrained in Ref.~\cite{GaQu09}. To analyze the sensitivity of the \elem{Be}{11} spectrum to the 3N force we modify the cutoff $\Lambda_{\text{3N}}$ or switch off the expected dominant contribution in the two-pion exchange term proportional to the low-energy constant (LEC) $c_{3}$ by setting the LEC to zero.
To constrain the LECs $c_D$ and $c_{E}$ that appear in the contact terms of the 3N force we follow the procedure described in Refs.~\cite{RoCa14,RoBi12}, where $c_{D}$ is fit to the triton-beta decay and $c_{E}$ to the ground-state energy of $A=3$ or $A=4$ systems. In particular, the NN+3N(400) interaction with the three-body cutoff $\Lambda_{\text{3N}}=400\,\text{MeV}$ is well established for applications within and beyond the mid-p shell~\cite{RoCa14,RoBi12,HeBi13,HeBo13,CiBa13,LaNa15}. 
 
In addition to this family of interactions we study a novel NN+3N interaction obtained at N$^2$LO, where the LECs of the NN and 3N contributions are fitted simultaneously to two-body scattering data and to ground-state binding energies and radii of $A=3$ and $4$ system,  \elem{C}{14}, \elem{O}{16} as well as binding energies of \elem{O}{22}, \elem{O}{24}, \elem{O}{25}~\cite{EkJa15}. 
The $\SAT$ interaction has been applied in several physics cases providing accurate predictions of the saturation properties of nuclear matter, nuclear binding energies~\cite{EkJa15} as well as observables sensitive to the neutron and charge distribution~\cite{HaEk16,MiBa16}.

The NN+3N interactions are softened via the similarity renormalization group (SRG)~\cite{Wegn94,BoFu07,SzPe00} at the three-body level up to a flow parameter of $\lambda_{\text{SRG}}=2.0\,\text{fm}^{-1}$ ($\alpha=0.0625\,\text{fm}^4$) as described in Refs.~\cite{RoLa11,RoCa14}. The SRG induced and initial 3N forces are treated explicitly at all steps of the calculation.
Due to the complexity of the calculations we use optimal parameters determined in previous studies~\cite{HuLa13, LaNa15}. 
The no-core shell model (NCSM) and NCSM with continuum calculations are carried out using a harmonic oscillator (HO) frequency of $\hbar\Omega=20\,\text{MeV}$ and an additional three-body truncation of the summed single-particle energies $e_1+e_2+e_3 \leq E_{3\text{max}}=14$.
To insure the convergence of our calculations we varied the HO model space $N_{\text{max}}$ and the number of NCSM eigenstates included in the NCSMC approach.
The technical details of the E1 transition calculations within the NCSMC approach are discussed in Ref.~\cite{NaQu16}.

The inclusion of the 3N force is computationally highly demanding and restricts the current application range of the NCSMC. 
Developments to combine the multi-reference normal-ordering~\cite{GeCa16} with the NCSMC approach are in progress, that will reduce the computational effort and make a variety of heavier exotic physical systems accessible.

\subsection{Additional reaction observables.}

\begin{figure}[t]
\includegraphics[clip, width=0.95\columnwidth]{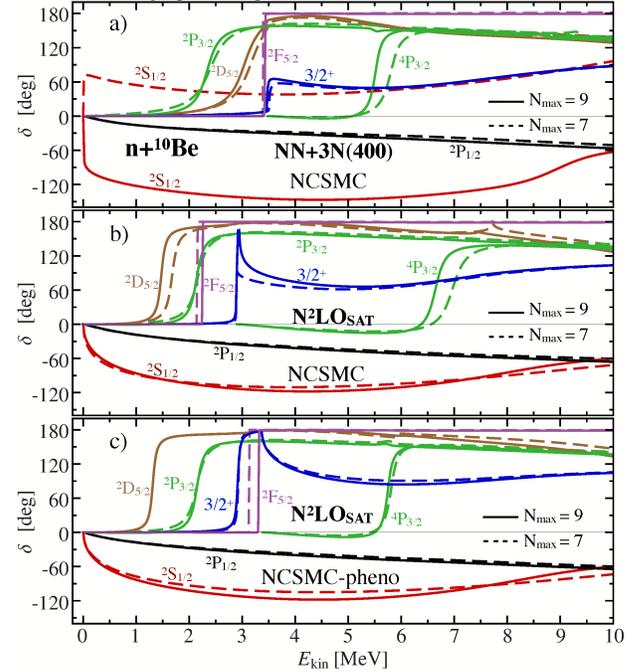}\\[-8pt] 
\caption{(color online) The n+\elem{Be}{10} phase shifts as function of the kinetic energy in the center-of-mass frame. 
a) NCSMC phase shifts for the NN+3N(400) and b) the $\SAT$ interaction, as well as c) NCSMC-pheno phase shifts for the $\SAT$ interaction. 
 }
\label{fig:Be11_PS_SUP}
\end{figure}

 In Fig.~\ref{fig:Be11_PS_SUP} we present a comparison of the n+\elem{Be}{10} phase shifts obtained from NCSMC calculations with and without phenomenological adjusted energies using the  NN+3N(400) or $\SAT$ interaction.
In all cases the comparison for $N_{\text{max}}=7$ and $9$ confirms the good convergence with respect to the model space. The NN+3N(400) interaction predicts a barely bound $1/2^+$ state that is a virtual state at $N_\text{max} = 7$ and explains the  strong change with $N_{\text{max}}$ for the $^{2}S_{1/2}$ phase shift. 

\begin{figure}[t]
\vspace{-4pt}
\hspace{-10pt}
\includegraphics[width=1.0\columnwidth]{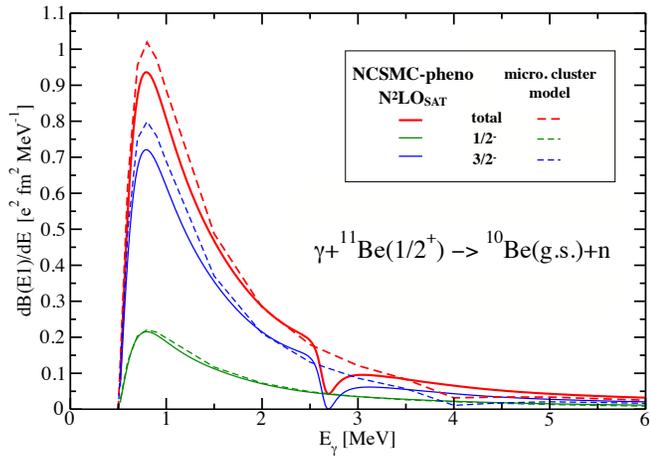}\\[-10pt]
\caption{(color online)
Dipole strength distribution $dB(E1)/dE$ of the photodisintegration process as
function of the photon energy. Partial wave contributions (thin lines)
to the full dipole strength distribution (thick lines) obtained with the
NCSMC-pheno using the $\SAT$ interaction (solid lines) compared to a
microscopic cluster model~\cite{Desc97} (dashed line).
}
\label{fig:Be11_photodis_SUP}
\end{figure}

The partial wave contributions to the dipole strength distribution obtained with the NCSMC-pheno using the $\SAT$ interaction is compared to a
microscopic cluster model~\cite{Desc97} in Fig.~\ref{fig:Be11_photodis_SUP}. 
The dip in the NCSMC-pheno calculations for the dipole strength distribution at about $2.7\,\text{MeV}$ can be also noticed, though much less pronounced, in microscopic cluster calculations around $4\,\text{MeV}$, which is close to the $3/2^-_1$ resonance energy ($3.6\,\text{MeV}$) obtained with that model~\cite{Desc97}.
%

\end{document}